\documentclass[twocolumn]{aastex631}
\usepackage[utf8]{inputenc}
\usepackage{natbib}
\usepackage{graphicx}
\usepackage{xcolor}
\usepackage{threeparttable}
\usepackage{multirow}

\received{}
\revised{}
\accepted{}
\submitjournal{ApJL}

\shorttitle{TRAPPIST-1{\rm c} with JWST NIRISS}
\shortauthors{Radica et al.}

\begin{document}

\title{Promise and Peril: Stellar Contamination and Strict Limits on the Atmosphere Composition of TRAPPIST-1c from JWST NIRISS Transmission Spectra}

\correspondingauthor{Michael Radica}
\email{radicamc@uchicago.edu}

\author[0000-0002-3328-1203]{Michael Radica}
\altaffiliation{NSERC Postdoctoral Fellow}
\affiliation{Department of Astronomy \& Astrophysics, University of Chicago, 5640 South Ellis Avenue, Chicago, IL 60637, USA}
\affiliation{Institut Trottier de Recherche sur les Exoplanètes and Département de Physique, Université de Montréal, 1375 Avenue Thérèse-Lavoie-Roux, Montréal, QC, H2V 0B3, Canada}

\author[0000-0002-2875-917X]{Caroline Piaulet-Ghorayeb}
\altaffiliation{E. Margaret Burbridge Postdoctoral Fellow}
\affiliation{Department of Astronomy \& Astrophysics, University of Chicago, 5640 South Ellis Avenue, Chicago, IL 60637, USA}
\affiliation{Institut Trottier de Recherche sur les Exoplanètes and Département de Physique, Université de Montréal, 1375 Avenue Thérèse-Lavoie-Roux, Montréal, QC, H2V 0B3, Canada}

\author[0000-0003-4844-9838]{Jake Taylor}
\affiliation{Department of Physics, University of Oxford, Parks Rd, Oxford OX1 3PU, UK}

\author[0000-0002-2195-735X]{Louis-Philippe Coulombe}
\affiliation{Institut Trottier de Recherche sur les Exoplanètes and Département de Physique, Université de Montréal, 1375 Avenue Thérèse-Lavoie-Roux, Montréal, QC, H2V 0B3, Canada}

\author[0000-0001-5578-1498]{Björn Benneke}
\affiliation{Institut Trottier de Recherche sur les Exoplanètes and Département de Physique, Université de Montréal, 1375 Avenue Thérèse-Lavoie-Roux, Montréal, QC, H2V 0B3, Canada}

\author[0000-0003-0475-9375]{Loic Albert}
\affiliation{Institut Trottier de Recherche sur les Exoplanètes and Département de Physique, Université de Montréal, 1375 Avenue Thérèse-Lavoie-Roux, Montréal, QC, H2V 0B3, Canada}

\author[0000-0003-3506-5667]{Étienne Artigau}
\affiliation{Institut Trottier de Recherche sur les Exoplanètes and Département de Physique, Université de Montréal, 1375 Avenue Thérèse-Lavoie-Roux, Montréal, QC, H2V 0B3, Canada}
\affiliation{Observatoire du Mont-Mégantic, Université de Montréal, Montréal, QC H3C 3J7, Canada}

\author[0000-0001-6129-5699]{Nicolas B.\ Cowan}
\affiliation{Department of Physics, McGill University, 3600 rue University, Montréal, QC, H3A 2T8, Canada}
\affiliation{Department of Earth and Planetary Sciences, McGill University, 3600 rue University, Montréal, QC, H3A 2T8, Canada}

\author[0000-0001-5485-4675]{René Doyon}
\affiliation{Institut Trottier de Recherche sur les Exoplanètes and Département de Physique, Université de Montréal, 1375 Avenue Thérèse-Lavoie-Roux, Montréal, QC, H2V 0B3, Canada}
\affiliation{Observatoire du Mont-Mégantic, Université de Montréal, Montréal, QC H3C 3J7, Canada}

\author[0000-0002-6780-4252]{David Lafrenière}
\affiliation{Institut Trottier de Recherche sur les Exoplanètes and Département de Physique, Université de Montréal, 1375 Avenue Thérèse-Lavoie-Roux, Montréal, QC, H2V 0B3, Canada}

\author[0009-0005-6135-6769]{Alexandrine L'Heureux}
\affiliation{Institut Trottier de Recherche sur les Exoplanètes and Département de Physique, Université de Montréal, 1375 Avenue Thérèse-Lavoie-Roux, Montréal, QC, H2V 0B3, Canada}

\author[0000-0003-4676-0622]{Olivia Lim}
\affiliation{Institut Trottier de Recherche sur les Exoplanètes and Département de Physique, Université de Montréal, 1375 Avenue Thérèse-Lavoie-Roux, Montréal, QC, H2V 0B3, Canada}

\begin{abstract}
Attempts to probe the atmospheres of rocky planets around M dwarfs present both promise and peril. While their favorable planet-to-star radius ratios enable searches for even thin secondary atmospheres, their high activity levels and high-energy outputs threaten atmosphere survival. Here, we present the 0.6--2.85\,$\mu$m transmission spectrum of the 1.1\,$\rm R_\oplus$, $\sim$340\,K rocky planet TRAPPIST-1\,c obtained over two JWST NIRISS/SOSS transit observations. Each of the two spectra displays 100--500\,ppm signatures of stellar contamination. Despite being separated by 367\,days, the retrieved spot and faculae properties are consistent between the two visits, resulting in nearly identical transmission spectra. Jointly retrieving for stellar contamination and a planetary atmosphere reveals that our spectrum can rule out hydrogen-dominated, $\lesssim$300$\times$ solar metallicity atmospheres with effective surface pressures down to 10\,mbar at the 3-$\sigma$ level. For high-mean molecular weight atmospheres, where O$_2$ or N$_2$ is the background gas, our spectrum disfavors partial pressures of more than $\sim$10\,mbar for H$_2$O, CO, NH$_3$ and CH$_4$ at the 2-$\sigma$ level. Similarly, under the assumption of a 100\% H$_2$O, NH$_3$, CO, or CH$_4$ atmosphere, our spectrum disfavors thick, $>$1\,bar atmospheres at the 2-$\sigma$ level. These non-detections of spectral features are in line with predictions that even heavier, CO$_2$-rich, atmospheres would be efficiently lost on TRAPPIST-1\,c given the cumulative high-energy irradiation experienced by the planet. Our results further stress the importance of robustly accounting for stellar contamination when analyzing JWST observations of exo-Earths around M dwarfs, as well as the need for high-fidelity stellar models to search for the potential signals of thin secondary atmospheres.
\end{abstract}


\section{Introduction} 
\label{sec: Introduction}

The detection and characterization of the atmospheres of rocky, Earth-sized planets is one of the key science goals of JWST. Not only is the presence of an atmosphere a top-level requirement for potential habitability \citep{Grenfell2020}, but constraining the composition of terrestrial planet atmospheres also has direct implications for our understanding of the formation and evolution of small planet atmospheres \citep{turbet_water_2023, krissansen-totton_predictions_2022, krissansen-totton_implications_2023, bolmont_water_2017, zahnle_cosmic_2017}.   

In recent years, the focus of such studies has been on small planets orbiting M dwarf stars, as the small size of late-type stars boosts the planet's atmosphere signal, either in transmission or emission, compared to Sun-like stars \citep[e.g.,][]{de_wit_combined_2016, wakeford_disentangling_2019, kreidberg_absence_2019, lim_atmospheric_2023, may_double_2023, greene_thermal_2023, zhang_gj_2024, kirk_jwstnircam_2024}. 

However, the potential promise of increased atmosphere observability also comes with a corresponding peril for atmosphere retention. Late-type stars have significantly longer pre-main sequence phases than Sun-like stars, during which time orbiting planets are subjected to large amounts of high-energy radiation \citep{Wordsworth2013, Wheatley2017}. Even at late times, during the M dwarf's main sequence phase, they still maintain large outputs of high energy radiation, with $L_X/L_{bol}$ ratios significantly higher than those of earlier-type stars \citep{Pizzolato2003, Wright2011, Wheatley2017, Lloyd2021}, reaching up to $\sim$2000$\times$ 
that of the Sun for ultra-cool M dwarfs like TRAPPIST-1 \citep{Wheatley2017, bourrier_reconnaissance_2017}. The increased amounts of high energy radiation received by M dwarf planets results in the potential for corresponding increases in the efficiency of atmosphere loss processes \citep[e.g.,][]{Wordsworth2013, Luger2015, bolmont_water_2017, turbet_water_2023}, resulting in uncertainty as to whether rocky planets around M dwarf stars can retain atmospheres to the present day \citep{zahnle_cosmic_2017}.

Moreover, late-type stars also maintain higher levels of activity and photospheric inhomogeneities (i.e., the presence of spots and faculae) than Sun-like stars \citep{Peacock2019, Lloyd2021}. Unocculted spots and faculae on the host star's photosphere (i.e., inhomogeneities that lie outside of a planet's transit chord), in particular, are a pernicious problem preventing the detection of atmospheres around Earth-like planets orbiting M dwarfs via transit observations \citep{rackham_transit_2018, rackham_transit_2019, wakeford_disentangling_2019, garcia_hst_2022, trappist-1_jwst_community_initiative_roadmap_2024, rackham_toward_2024}. The presence of unocculted inhomogeneities can impart spurious features on the transmission spectra of rocky planets, which can often be significantly larger than expected atmosphere features \citep{rackham_transit_2018, rackham_effect_2023, lim_atmospheric_2023, trappist-1_jwst_community_initiative_roadmap_2024, rackham_toward_2024}. This phenomenon is often referred to as the transit light source effect (TLSE). Despite these challenges, transit observations using the Hubble Space Telescope (HST) and now with JWST, have been able to widely rule out cloud-free, H/He-dominated atmospheres for rocky M dwarf planets \citep{de_wit_combined_2016, wakeford_disentangling_2019, garcia_hst_2022, Libby-Roberts2022, lim_atmospheric_2023, lustig-yaeger_jwst_2023, cadieux_transmission_2024, Damiano2024, kirk_jwstnircam_2024}, although such extended atmospheres are often already disfavoured based on mass and radius measurements alone \citep[e.g.,][]{agol_refining_2021}. In many cases though, the TLSE, still prevents us from robustly probing secondary, or high-mean molecular weight atmospheres in transmission \citep[e.g.,][]{wakeford_disentangling_2019, lim_atmospheric_2023, cadieux_transmission_2024, may_double_2023, moran_high_2023}. 

Observations of M dwarf planets in eclipse geometry circumvent the TLSE problem but bring with them their own sets of challenges: the planet's orbital eccentricity must be well known to calculate the phase of the eclipse, emission observations favour planets with high temperatures potentially ruling out habitable-zone worlds, etc. Though, even without the impacts of the TLSE, emission studies of rocky M-dwarf planets, both in the JWST era and previously, have either been negative (\citealp[i.e., favour atmosphereless interpretations;][]{kreidberg_absence_2019, greene_thermal_2023, ih_constraining_2023, zhang_gj_2024, mansfield_no_2024, xue_jwst_2024}) or unable to provide definitive evidence for the presence of a high-mean molecular weight (MMW) atmosphere \citep[e.g.,][]{zieba_no_2023, lincowski_potential_2023}, often due to degeneracies between atmospheric absorption and surface reflectivity \citep[e.g.,][]{ih_constraining_2023, lincowski_potential_2023}. 

Here, we present transit observations of TRAPPIST-1\,c with JWST NIRISS/SOSS. TRAPPIST-1\,c is the second innermost planet ($a$=0.01580\,AU, $T_{eq}$=340\,K) of the TRAPPIST-1 system, with a mass and radius of 1.3$\times$ and 1.1$\times$ that of the Earth, respectively \citep{gillon_temperate_2016, Ducrot2020, agol_refining_2021}. It was first observed in transit by \citet{de_wit_combined_2016} with HST/WFC3 from 1.1 -- 1.7\,µm. Their observations rule out cloud-free, H/He-dominated atmosphere scenarios at $>$10-$\sigma$, but lack the precision to constrain the presence of a wide range (e.g., H$_2$O-rich, Venus-like) of higher-MMW atmospheres. 

TRAPPIST-1\,c was then observed in emission with JWST MIRI/F1500W photometry by \citet{zieba_no_2023} at 15\,µm. Their dayside brightness temperature of 380$\pm$31\,K rules out thick ($>$10\,bar), CO$_2$-rich atmospheres, but remains consistent with a thin (0.01\,bar) atmosphere composed purely of CO$_2$, or trace amounts of CO$_2$ in thinner (0.1 -- 10 \,bar), O$_2$-dominated atmospheres at a $\sim$2$\sigma$ level. This was then followed up by \citet{lincowski_potential_2023} who compared the \citet{zieba_no_2023} eclipse depth to a broader range of potential atmosphere scenarios using a self-consistent modelling framework. Their results generally agree with those of \citet{zieba_no_2023}, in that trace amounts of CO$_2$ in a thin O$_2$-dominated atmosphere provide the best matches to the data. However, they also also find that $\sim$0.1\,bar pure-O$_2$, and even $\lesssim$3\,bar steam atmospheres are consistent with the data within $\lesssim$2$\sigma$. Venus-like atmospheres, though, are disfavoured at $\gtrsim$3$\sigma$. 

This paper is organized as follows. We present the observations and data analysis in Section~\ref{sec: Observations}. We describe our analysis of the stellar photosphere in Section~\ref{sec: TLSE Modelling}, and joint modelling of the atmosphere and stellar contamination in Section~\ref{sec: Atmosphere Modelling}. We then discuss our results in Section~\ref{sec: Discussion} and follow up with a short conclusion in Section~\ref{sec: Conclusions}.

\section{Observations \& Data Analysis} 
\label{sec: Observations}

We observed two transits of TRAPPIST-1\,c with the SOSS mode of JWST's NIRISS instrument \citep{albert_near_2023, doyon_near_2023} as part of program GO 2589 (PI: O.\ Lim). The first visit started at 18:37:51 UTC on October 28, 2022, and the second at 21:09:46 UTC on October 31, 2023. Each visit lasted a total of 4.6\,hr, and consisted of 159 integrations with 18 groups per integration.

\subsection{Data Reduction}
\label{sec: Reduction}
We reduce the time series observations (TSO) with the \texttt{exoTEDRF} pipeline\footnote{\url{https://github.com/radicamc/exoTEDRF}} \citep{radica_exotedrf_2024, feinstein_early_2023, radica_awesome_2023}, closely following standard procedures as described in, e.g., \citet{radica_muted_2024, cadieux_transmission_2024, piaulet-ghorayeb_jwst_2024}. Concretely, we perform the standard \texttt{exoTEDRF} stage 1 calibrations, including the correction of 1/$f$ noise at the group-level (that is, before ramp fitting) using the \texttt{scale-achromatic-window} method introduced in \texttt{exoTEDRF~v1.4.0}. With this algorithm, we only use rows within 30 pixels of the trace to estimate the 1/$f$ level, which can potentially help to reduce residual red noise in SOSS TSOs, particularly with high group numbers \citep[e.g.,][]{feinstein_early_2023, holmberg_exoplanet_2023}. We then proceed with the standard Stage 2 calibrations (e.g., flat field correction, bad pixel interpolation, etc.). Notably, we perform a ``piecewise'' background subtraction \citep[e.g.,][]{lim_atmospheric_2023, fournier-tondreau_near-infrared_2024, radica_muted_2024}, whereby the standard STScI SOSS background model is scaled separately on either side of the background ``step'', as we found that a single scaling for the whole detector did not adequately remove the background signal. As a final step in the calibrations, we performed a principle component analysis (PCA) on the 2D detector images \citep[e.g.,][]{coulombe_broadband_2023, radica_muted_2024} to identify any detector-level trends (i.e., mirror tilt events, trace position drifts), which could impact the light curves.

We extract the stellar spectrum using a simple box aperture with a width of 40 pixels, as the dilution effects due to the SOSS first and second order overlap are expected to be negligible for transit measurements \citep{darveau-bernier_atoca_2022, radica_applesoss_2022}. Finally, there are several undispersed contaminants due to background field stars (so-called ``order 0'' contaminants) present on the detector for both visits, several of which partially intersect the target spectral traces. These contaminants are located at wavelengths 0.670--0.675\,µm, 1.642--1.658\,µm, and 2.064--2.072\,µm, for visit 1, and 0.975--1.001\,µm, 1.251--1.279\,µm, and 1.567--1.578\,µm for visit 2. We simply mask these contaminants in the extracted spectra.

\subsection{White Light Curve Fitting}
\label{sec: White Light Fitting}

\begin{figure*} 
	\centering
	\includegraphics[width=\textwidth]{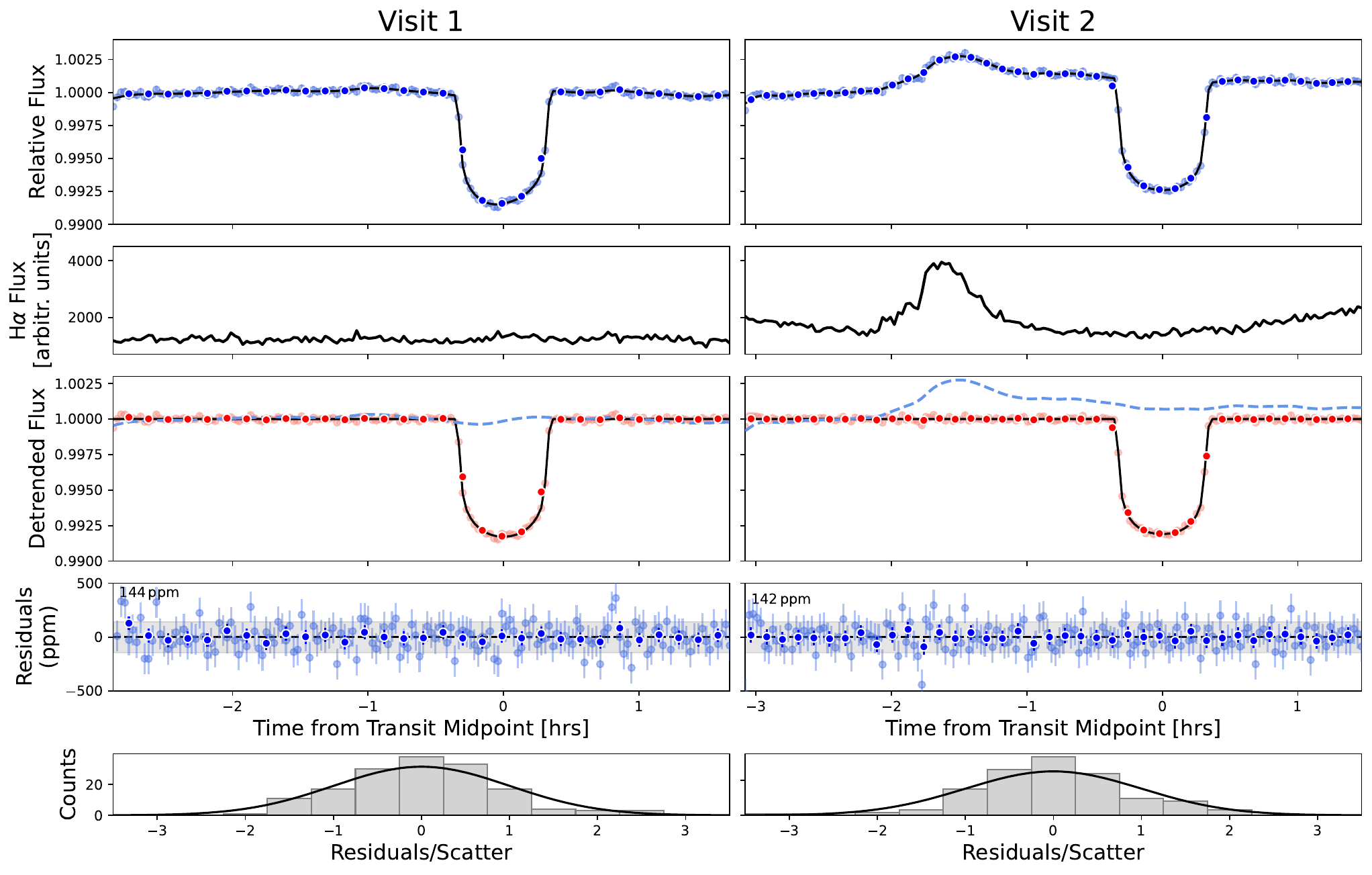}
    \caption{White light curves and fit results for both visits.
    \emph{Top row:} Raw white light curves for order 1, with the best-fitting astrophysical + systematics model overplotted in black. A flare is visible in the pre-transit baseline for visit 2. 
    \emph{Second row:} H$\alpha$ time series (in arbitrary units) for each visit. The time series is relatively quiescent in visit 1, but traces the flare structure in visit 2.
    \emph{Third row}: White light curves after removal of the systematics model described in Section~\ref{sec: Observations}. The best-fitting astrophysical transit model is overplotted in black, and the systematics model that has been removed from the data is shown in the blue dashed line. 
    \emph{Fourth row}: Residuals after the removal of the astrophysical and systematics models.
    \emph{Bottom row}: Histogram of residuals. 
    \label{fig:WhiteLight}}
\end{figure*}

We construct white light curves for each visit and each order by summing all flux on the detector in order 1 (0.85 -- $\sim$2.8\,µm), and only wavelengths $<$0.85\,µm in order 2. We fit these white light curves using the flexible \texttt{exoUPRF} library \citep{radica_exouprf_2024}. For each visit, we jointly fit the first and second order white light curves, as in \citet{radica_muted_2024}. Concretely, we fit an astrophysical transit model, as computed by \texttt{BATMAN} \citep{kreidberg_batman_2015}, with the orbital parameters ($T_0$, the time of mid-transit, $inc$, the orbital inclination, and $a/R_*$, the scaled orbital semi-major axis) shared between both orders, and the scaled planet radius, $R_p/R_*$ fit to each order individually. We, furthermore, assume a circular orbit and fix the orbital period to 2.421937\,d \citep{agol_refining_2021}. We fit, separately for each order, the two parameters of the quadratic limb darkening law, $u_1, u_2 \in [-1,1]$, in order to avoid potential biases that can be introduced by employing the \citet{kipping_efficient_2013}, parameterization \citep[e.g.,][]{Coulombe2024_biases}. We also tested several other limb-darkening treatments but found that the choice of parameterization did not impact the final spectra.

We also fit a systematics model to the light curve for each order. For each visit, the first component of the systematics model is a linear trend with time. TRAPPIST-1 is well-known to be a highly active star \citep{lim_atmospheric_2023}, and we, therefore, include a Gaussian process (GP) with a Matérn 3/2 kernel, to model residual red noise in the visit 1 light curves not captured by the aforementioned detrending. We train the GP on the timeseries of the second eigenvalue extracted from the PCA, which roughly corresponds to sub-pixel drifts of the trace position on the detector, as we find this reduces the residual scatter in the light curves by a further $\sim$10\% compared to training the GP on the light curve time axis. The characteristic amplitude of the GP was fit independently for each order, but the timescale was shared between both \citep[e.g.,][]{radica_muted_2024}.

During the pre-transit baseline of the second visit, there was a large flare event on the host star. We to correct this flare in the light curves by detrending against the H$\alpha$ flux, computed for each integration by measuring the area between the flux in the H$\alpha$ line and a local quadratic continuum, shown in the second panels in Figure~\ref{fig:WhiteLight}. However, this avenue was unsuccessful as there is a slight time delay between the appearance of the flare in the H$\alpha$ time series and in the white light curve, likely due to lags in energy release as a function of wavelength \citep{howard_characterizing_2023}. However, given the broad structure of this flare compared to others seen in TRAPPIST-1 light curves \citep[e.g.,][]{howard_characterizing_2023}, we found that we were able to adequately include the flare in our light curve fits using a GP which again employed the Matérn 3/2 kernel. As above, we fit the characteristic amplitude separately to each order, but shared the timescale between both.

Finally, we include an error inflation term for each order of each visit, which is added in quadrature to the extracted flux errors. Our final models consist of 20 parameters for each visit --- we use wide and uninformative priors in all cases. The results of the white light curve fits are shown in Figure~\ref{fig:WhiteLight} and the best-fitting astrophysical parameters are summarized in Table~\ref{tab: WLC Parameters}. 

\subsection{Spectroscopic Light Curve Fitting}
\label{sec: Spec Light Fitting}

To construct the spectrophotometric light curves, we sum the first order flux in bins of 80 pixels (corresponding to a constant resolving power of approximately $R$=25). The combined effects of the spectrum of the late-M host star, and the low throughput of the SOSS order 2 mean that the second order trace is very faint for observations of TRAPPIST-1 planets \citep[e.g.,][]{lim_atmospheric_2023}. We therefore do not bin order 2 further, and simply include the white light curve results in our transit spectrum. The orbital parameters from both visits are consistent between visits (Table~\ref{tab: WLC Parameters}) as well as with literature values \citep[e.g.,][]{agol_refining_2021}, and so we fix them to the weighted average between the two visits when fitting the spectroscopically binned light curves. We include the same systematics model as mentioned above for each visit, except that we fix the characteristic timescale of the GP to the best-fitting value from the white light fit for each visit, and allow the amplitude to freely vary, since we can reasonably expect the impacts of stellar variability to be more prominent at bluer vs.\ redder wavelengths \citep[e.g.,][]{radica_muted_2024}.

We also reduce and fit the TSOs from each visit using the independent \texttt{NAMELESS} pipeline \citep{feinstein_early_2023, coulombe_broadband_2023}. Further details of this analysis are presented in Appendix~\ref{sec: Additional Reductions}, and a comparison of the spectra from both pipelines are shown in Figure~\ref{fig:CompareSpectra}. Since the spectra from each pipeline are consistent to within 1-$\sigma$ for both visits (or $\lesssim$1 via a $\chi^2$-per-data-point metric), we only consider the \texttt{exoTEDRF} spectrum for the remainder of our analysis.

\section{Modelling of the Stellar Photosphere}
\label{sec: TLSE Modelling}

Before undertaking any planet atmosphere analysis, we first seek to better constrain the properties of TRAPPIST-1's photosphere, and thereby aid in our attempts to disentangle veritable atmosphere signals from the TLSE. To this end, we perform two independent, but complementary analyses: one of the in-transit data via the TLSE (Section~\ref{sec: in transit}), and another focusing on the out-of-transit data via modelling the stellar spectrum itself (Section~\ref{sec: out of transit}).

\subsection{Transit Light Source Effect Modelling}
\label{sec: in transit}

\begin{figure*} 
	\centering
	\includegraphics[width=0.8\textwidth]{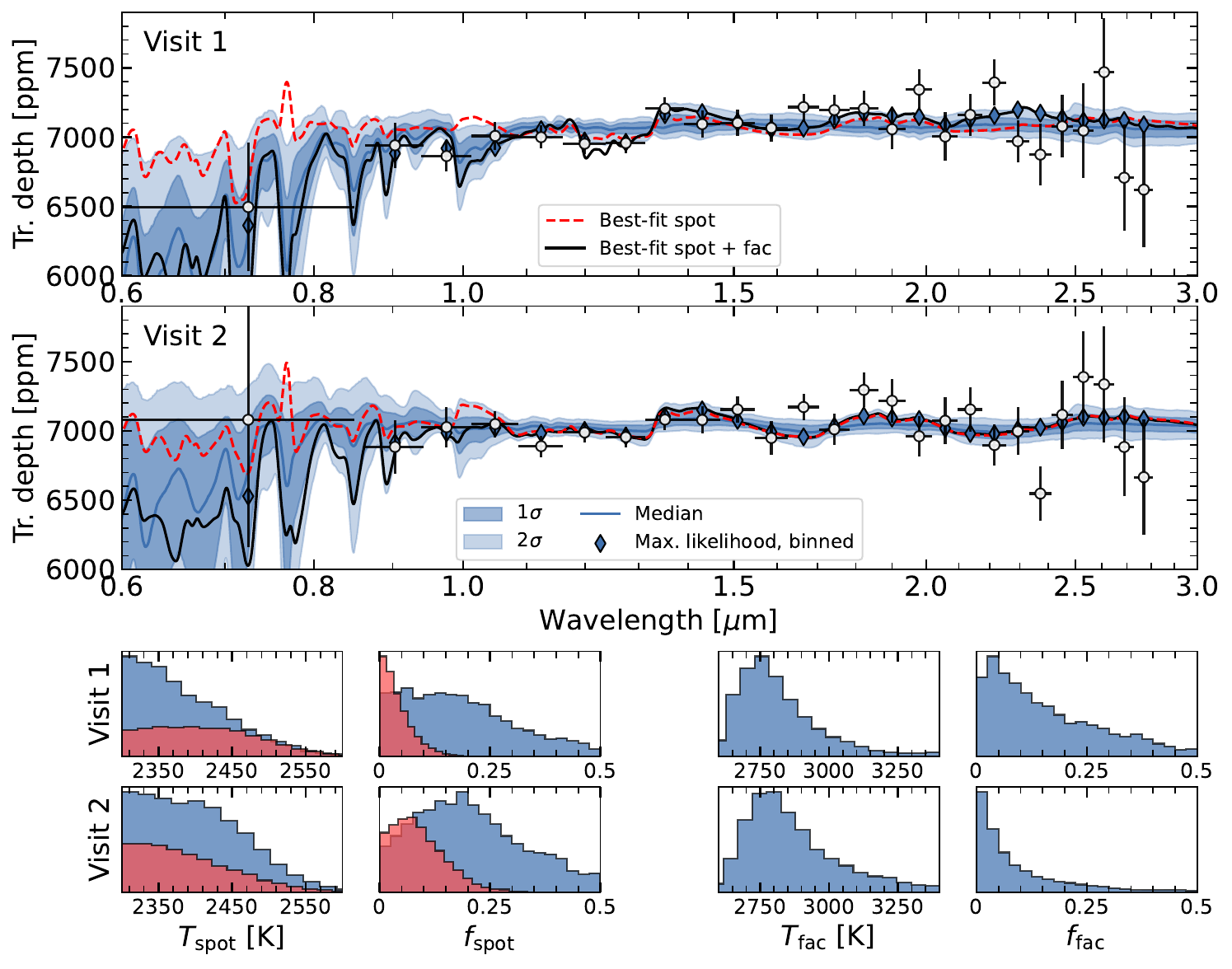}
    \caption{Results from the stellar-contamination-only fits to the transmission spectrum of each visit of TRAPPIST-1\,c. \emph{Top panel:} Best-fitting unocculted heterogeneity model (black line) to the observed visit 1 spectrum (black points). The median, 1-, and 2-$\sigma$ ranges from the posterior spectra (colour shading, smoothed to a resolving power of 100) are also shown, integrated within each of the spectral wavelength bins (diamonds), for easier comparison. We assume a contribution from both spots and faculae and that the photosphere component represents the transit chord. The best-fitting model obtained from a retrieval where only spots are considered is shown in dashed red.
    \emph{Second panel:} Same as top, but for the fit to the visit 2 spectrum. 
    \emph{Bottom panels:} Posterior probability distributions of the spot and faculae temperatures, $T_\mathrm{spot}$ and $T_\mathrm{fac}$, and covering fractions, $f_\mathrm{spot}$ and $f_\mathrm{fac}$ for both visits. The corresponding distributions of spot properties from the spot-only retrieval are overlaid in red}. Stellar contamination alone can explain the wavelength-dependent variations in the spectrum, and the retrieved heterogeneity parameters are largely consistent across both visits.
    \label{fig:stellar_contamination_results}
\end{figure*}

We first model the transmission spectrum of TRAPPIST-1\,c assuming that the TLSE can completely explain each visit's transmission spectrum. This approach is motivated by the presence of small features resembling water absorption (e.g., around 1.4\,µm; Figure~\ref{fig:stellar_contamination_results}) and a pronounced slope towards short wavelengths in the spectrum of the first visit. We use the open-source code \texttt{stctm} \citep{stctm,piaulet-ghorayeb_jwst_2024} to perform the TLS retrievals. We model the presence of two populations of heterogeneities: spots (covering a fraction $f_\mathrm{spot}$ of the photosphere, and cooler than the photosphere at a temperature $T_{\rm{spot}}<T_{\rm{phot}}$), and faculae (with $f_\mathrm{fac}$ representing their covering fraction, and $T_{\rm{fac}}>T_{\rm{phot}}$ their temperature).

For each population of heterogeneities, we fit for the temperature difference to the photosphere ($T_\mathrm{het} = T_\mathrm{phot} + \Delta T_\mathrm{het}$), and place a conservative uniform prior from 0 to 50\% on the covering fraction for each of the components, $f_\mathrm{het}$. The stellar photosphere temperature is fitted using a Gaussian prior, with the mean set by the literature value of 2566\,K \citep{agol_refining_2021} and a standard deviation of 50\,K. We allow faculae to be at most 1000\,K hotter than the photosphere (with an otherwise uniform prior on $\Delta T_\mathrm{fac}>0$), while the minimum temperature of spots is set by the minimum temperature for which stellar models are available in our grid (2300\,K). In the retrieval, we select at each iteration the closest-matching model (in effective temperature and surface gravity space) for the photosphere, spot, and faculae components from a finely-sampled pre-computed grid obtained with the \texttt{MSG} module \citep{townsend_msg_2023} using the PHOENIX \citep{husser_new_2013} stellar model grid. We also introduce the fitting of $\Delta \log g_\mathrm{het}$ (Piaulet-Ghorayeb et al., in prep), the $\log g$ difference between the photosphere and heterogeneity components, allowing the heterogeneity spectra to use a lower $\log g_\mathrm{het}$ than that of the photosphere, motivated by recent work \citep[e.g.,][]{fournier-tondreau_near-infrared_2024}.

We use the affine-invariant Markov Chain Monte Carlo sampler \texttt{emcee} \citep{foreman-mackey_emcee_2013} to explore the parameter space, with the number of walkers set as 20 times the number of fitted parameters, and run the chains for 5000 steps, visually monitoring for convergence, and discard  the first 60\% as burn-in. In addition to the stellar heterogeneity properties, we fit the wavelength-independent transit depth of TRAPPIST-1\,c, $D$. We perform independent fits to the spectrum of each visit, as they are separated by about one year and should not, a priori, be affected by the same TLS contribution. We also perform additional retrievals considering only faculae or only spots in addition to the quiet stellar photosphere.

We find that when either only spots or both spots and faculae are included in our model, we obtain a very good match to the observed spectra, without the need for a planetary atmosphere. The addition of faculae enables us to fit better the shortest-wavelength data points with a tentative downward slope which cannot be created by spots alone (Figure \ref{fig:stellar_contamination_results}) and results in broader distributions on the spot properties given the interplay with a facula component. However, when only faculae are included in our retrievals, the resulting TLS model does not capture the observed spectral features, particularly the broad absorption features around 1.4 and 1.9\,µm (as faculae would predict these features to instead be inverted). The resulting parameters from these retrievals are therefore not presented, as they result in poorer fits. 

We find that similar spot and facula properties can explain both visits (Figure \ref{fig:stellar_contamination_results}; Table~\ref{tab: Stellar Parameters}), which motivates the use of a shared stellar contamination component in our joint retrievals of stellar and planetary atmosphere contributions to the spectra in Section~\ref{sec: Atmosphere Modelling}. There is, however, a slight ($<1\sigma$) offset between the ``bare-rock'' transit depths retrieved in the fit to visit 1 ($D_\mathrm{v1} = 7035^{+152}_{-170}$ ppm) and visit 2 ($D_\mathrm{v2} = 6859^{+141}_{-136}$ ppm). Finally, our retrievals do not support the use of different $\log g$ values for the photosphere and heterogeneity components, as we only find lower limits on $\Delta \log g_\mathrm{het}$ (e.g., Figure~\ref{fig:TLS_inferences}).

We use these TLS-only retrievals to produce ``TLS-corrected'' spectra which we use to explore the limiting effect of stellar contamination on the constraining power of our observations (see Section~\ref{sec: TLS Limitations}; Figure~\ref{fig:TLS_marg}), and for visualization purposes in order to interpret the results from our atmosphere retrievals (Figures \ref{fig:met_psurf} and \ref{fig:atmosphere_molbymol}). For each visit, the TLS-corrected spectrum is produced by dividing the observed spectrum by the best-fitting stellar contamination model, and then multiplying the result by the best-fit model bare-rock transit depth (6776 ppm for visit 1, 6730 ppm for visit 2). These spectra aim to quantify the precision we could reach on atmospheric inferences given our observing setup if stellar contamination was not a factor.

Finally, to obtain the joint visit 1+2 TLS-corrected spectrum shown on Figures \ref{fig:met_psurf} and \ref{fig:atmosphere_molbymol} (but not used in the retrievals), we simply offset the visit 2 spectrum by the difference between the best-fit model bare-rock transit depths of each visit, and then average the two spectra.

\subsection{Out-of-Transit Stellar Spectrum Modelling}
\label{sec: out of transit}

In order to verify the reliability of our TLS retrievals described above, as well as gain additional insights into the heterogeneous nature of the photosphere of TRAPPIST-1, we directly model the out-of-transit stellar spectrum, roughly following the methodologies of \citet{wakeford_disentangling_2019} and \citet{moran_high_2023}. For each visit, we first create a median stack of the stellar spectrum considering only the out-of-transit integrations. For visit 1, this includes both the pre- and post-transit baseline. For visit 2, we only include integrations before the flare event in the pre-transit baseline. The post-transit baseline appears, by eye, to be at a slightly higher level than the pre-transit baseline. We test fits for visit 2 using only the pre-transit and only the post-transit baselines, as well as both and note that we do not find any difference in the inferred photosphere or heterogeneity parameters as a result of these choices, and we therefore proceed using both the pre- and pos-transit baseline for the fits. 

We then flux calibrate the stellar spectra, following the procedure presented in \citet{lim_atmospheric_2023}. Like \citet{moran_high_2023}, we use the standard deviation of the stellar spectra as a function of time as the errors on each wavelength point, as we find this more representative than the pipeline-extracted flux errors. 

As with the TLS fits, we proceed to fit one-, two-, and three-component models (i.e., photosphere-only, photosphere + spots/facule, photosphere + spots + faculae) directly to the stellar spectra using the \texttt{StellarFit}\footnote{\url{https://github.com/radicamc/StellarFit}} package. For consistency with the in-transit fits, we explore a grid of PHOENIX model spectra \citep{husser_new_2013}, but also test a grid of SPHINX models \citep{iyer_sphinx_2023}, which include temperatures as low as 2000\,K, more appropriate for modelling cold spots on ultra-cool dwarfs like TRAPPIST-1. All models in a given grid are scaled by $R_*^2/D^2$ to represent the flux received at Earth, using $R_*$=0.1192\,$R_\odot$ for the stellar radius \citep{agol_refining_2021} and $D$=12.1\,pc for the distance of the system to the Earth \citep{gillon_temperate_2016}. 

\begin{figure*} 
	\centering
	\includegraphics[width=0.8\textwidth]{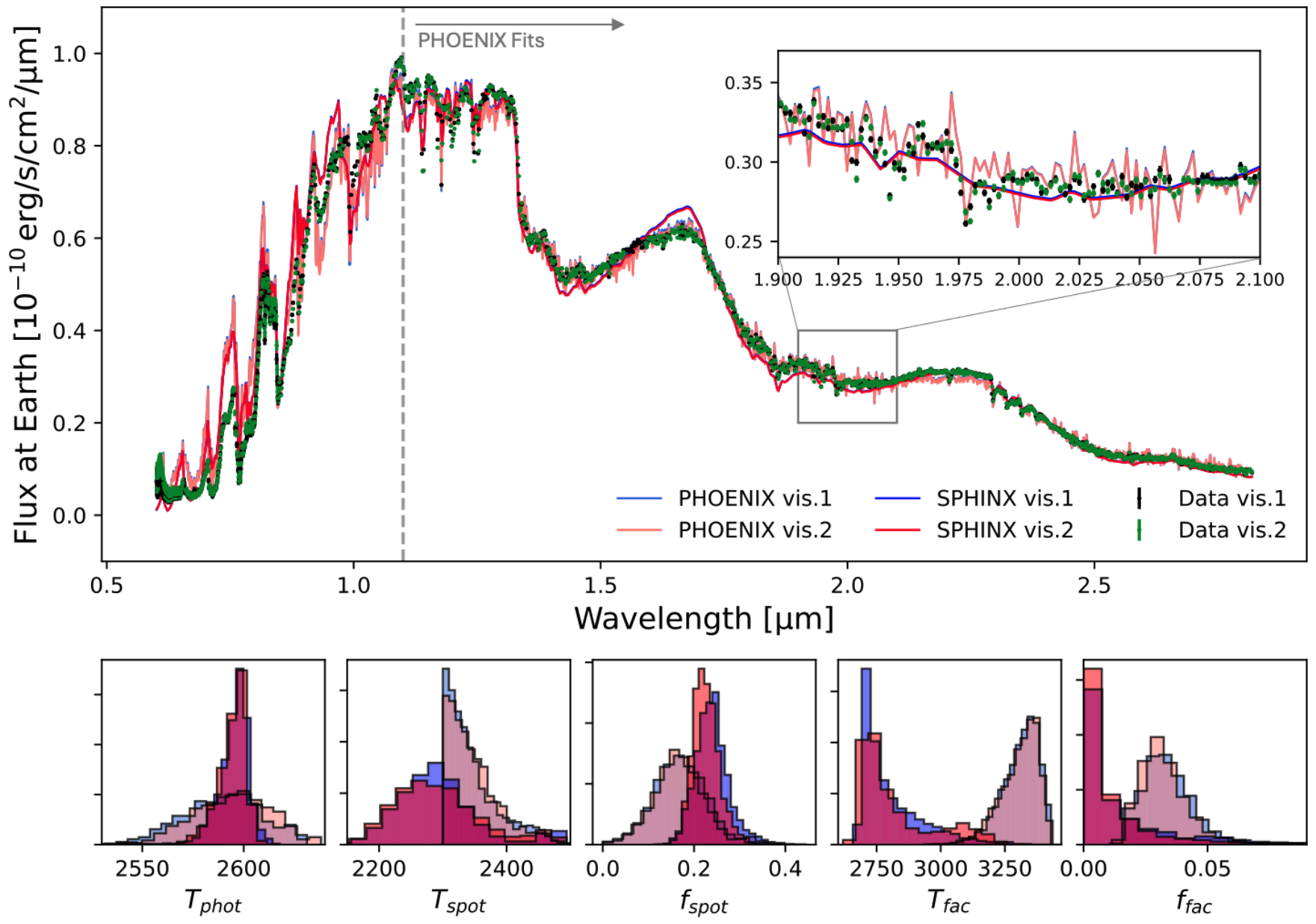}
    \caption{Results from the out-of-transit stellar spectrum fits.
    \emph{Top}: Flux calibrated stellar spectra for visit 1 (black) and visit 2 (green) along with best-fitting PHOENIX and SPHINX three-component (i.e., photosphere + spots + faculae) stellar models. Visit 1 models are shown in blue and visit 2 in red, whereas PHOENIX models are denoted by lighter colours and SPHINX models by darker colours. PHOENIX models are only fit to wavelengths $>$1.1\,µm as the fits to the shortest wavelengths were exceptionally poor. A zoom-in to a region of the spectrum is also included to visualize fine structure.
    \emph{Bottom}: Posterior distributions for the photosphere temperature, as well as spot and faculae temperatures and distributions. As above, visit 1 is shown in blue and visit 2 in red, whereas darker colours indicate SPHINX models and lighter colours PHOENIX models. Note that the cutoff in the spot temperature at 2300\,K for the PHOENIX grid is due to models not being available at colder temperatures.
    \label{fig:stellar_spectrum_results}}
\end{figure*}

One-component fits have four free parameters: the stellar effective temperature, $T_{phot}$, the surface gravity, as well as a spectrum scaling factor \citep[e.g.,][]{wakeford_disentangling_2019, moran_high_2023}, and a multiplicative error inflation term. The photosphere temperature was allowed to vary from 2300 -- 5000\,K for the PHOENIX grid or 2000 -- 4000\,K for the SPHINX grid, the log gravity from 3.5 -- 5.5 (PHOENIX) or 4.0 -- 5.5 (SPHINX), and the scale factor from 0.8 -- 1.2. Multi-component fits also include the spot/faculae temperature ($T_\mathrm{spot}$/$T_{fac}$), covering fraction ($f_\mathrm{spot}$/$f_\mathrm{fac}$), and the gravity of the heterogeneity \citep[e.g.,][]{lim_atmospheric_2023, fournier-tondreau_near-infrared_2024}. Spot/faculae temperatures are required to be at least 100\,K and up to 1000\,K cooler/warmer than the photosphere, whereas their gravity must be within 1.0 of the photosphere value. 

Unlike previous works, we do not simply consider a ``best-fitting'' model, but instead explore the full parameter space of model fits to obtain distributions on the photosphere and heterogeneity parameters analogous to our TLS retrievals. We use nested sampling for the posterior exploration, as implemented in the \texttt{dynesty} \citep{speagle_dynesty_2020} package, which also allows us to assess the statistical evidence for including heterogeneities in our models. The best fitting three-component models for both stellar grids are shown in Figure~\ref{fig:stellar_spectrum_results}, and are in excellent agreement with the in-transit TLS results. The best-fitting parameters from both the in-transit and out-of-transit analyses are also presented in Table~\ref{tab: Stellar Parameters}.

\section{Joint modelling of atmosphere and stellar contamination} 
\label{sec: Atmosphere Modelling}

\begin{figure*}
    \centering
    \includegraphics[width=\linewidth]{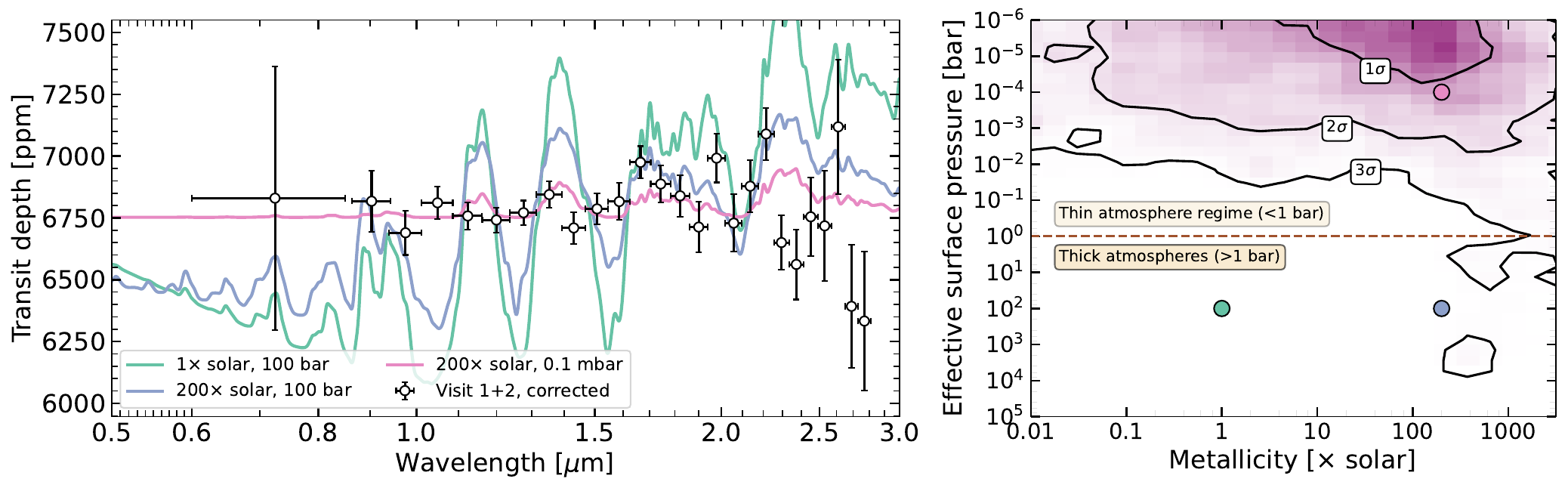}
    \caption{Constraints on H$_2$/He-dominated atmospheres on TRAPPIST-1\,c from joint atmosphere and TLSE modelling. 
    \textit{Left panel:} Joint visit 1+2 spectrum, corrected by the best-fitting stellar contamination model (black points), with three representative atmosphere forward models overplotted. 
    \textit{Right panel:} Joint posterior probability on the effective surface pressure of a H$_2$/He-dominated atmosphere and its metallicity from the SCARLET atmosphere retrievals (colour shading). We show the 1, 2, and 3$\sigma$ confidence regions and three markers corresponding to the three models shown in the left panel. At low metallicities, we obtain a strict 3-$\sigma$ upper limit of $\sim$10\,mbar on the effective surface pressure of any H$_2$/He-dominated atmosphere, although thick high-metallicity atmospheres ($>$100$\times$ solar) are possible within the 3$\sigma$ contours.
    \label{fig:met_psurf}}
\end{figure*}

We follow up our analysis of stellar contamination by using SCARLET \citep{benneke_atmospheric_2012, benneke_how_2013, benneke_strict_2015,benneke_sub-neptune_2019, benneke_water_2019, pelletier_where_2021, piaulet_evidence_2023} to more thoroughly explore the potential atmosphere scenarios consistent with our observation, while still accounting for the contributions of stellar contamination. The forward modelling component of SCARLET iteratively solves the radiative transfer equation and the 1D vertical structure of the atmosphere under the assumption of hydrostatic equilibrium, given the vertical temperature-pressure profile and the molecular abundances in each atmospheric layer. In the retrieval, the planetary radius is optimized for each set of sampled parameters to obtain the best match to the observed spectrum. 

We perform three categories of retrievals corresponding to different assumptions about the planetary atmosphere: (1) H$_2$/He-dominated scenarios; (2) single-component high-MMW atmospheres; (3) two-component high-MMW atmospheres with one infrared absorber present in the atmosphere and either N$_2$ or O$_2$ acting as the background gas.  

We parameterize H$_2$/He-dominated atmospheres in the retrieval with a H$_2$/He mixture as the background gas (assuming He/H$_2$= 0.157), log-uniform priors on the abundances of N$_2$, CH$_4$, H$_2$O, CO, CO$_2$, and NH$_3$, and a log-uniform prior on the pressure at which the atmosphere becomes opaque (either due to the presence of a cloud deck or a solid surface). We use the HELIOS-K \citep{heliosK_2015,heliosK2_2021} compiled line lists for CH$_4$ \citep{hargreaves_accurate_2020}, H$_2$O \citep{ExoMol_H2O}, CO \citep{Hargreaves2019}, CO$_2$ \citep{yurchenko_exomol_2020}, and NH$_3$ \citep{coles_exomol_2019}.

For single-component high-MMW atmospheres, we only explore absorbers with prominent features over the NIRISS/SOSS wavelength range; namely, atmospheres made of 100\% CH$_4$, H$_2$O, CO, or NH$_3$. We also fit for the ``effective surface pressure'' at which the atmosphere becomes opaque. 

Finally, we also explore potential N$_2$-dominated atmospheres (similar to some terrestrial objects in the solar system), or O$_2$-dominated atmospheres (motivated by the possible O$_2$-CO$_2$ atmosphere consistent with the MIRI eclipse observations of TRAPPIST-1\,c; \citealp{zieba_no_2023}) where only one of CH$_4$, H$_2$O, CO, or NH$_3$ acts as an additional absorber beyond the broad N$_2$-N$_2$ or O$_2$-O$_2$ collision-induced absorption features. In this retrieval setup, we fit for the partial pressure of each component. Unfortunately, the CH$_4$ line lists does not extend all the way to the short-wavelength end of NIRISS/SOSS order 2 \citep{hargreaves_accurate_2020}. Therefore, we exclude the order 2 data points from the spectrum for the 100\% CH$_4$, N$_2$-CH$_4$, and O$_2$-CH$_4$ retrievals.

We assume that the vertical temperature structure is isothermal, motivated by the fact that our transmission spectra would not the precision to measure any temperature gradient in the thin transit photosphere region we are probing. We use the nested sampling method \citep{skilling_nested_2004, skilling_nested_2006} to sample the full parameter space, with the multi-ellipsoid method implemented using the \texttt{nestle} module\footnote{\url{https://github.com/kbarbary/nestle}} within the SCARLET retrieval framework. The final posterior distributions are obtained from retrievals performed using at least 1000 live points. We compute forward models at R=31,250 before convolving them to the resolution of the spectrum, assuming a uniform throughput within each bin for the likelihood calculation.

\begin{figure*}
    \centering
    \includegraphics[width=0.7\linewidth]{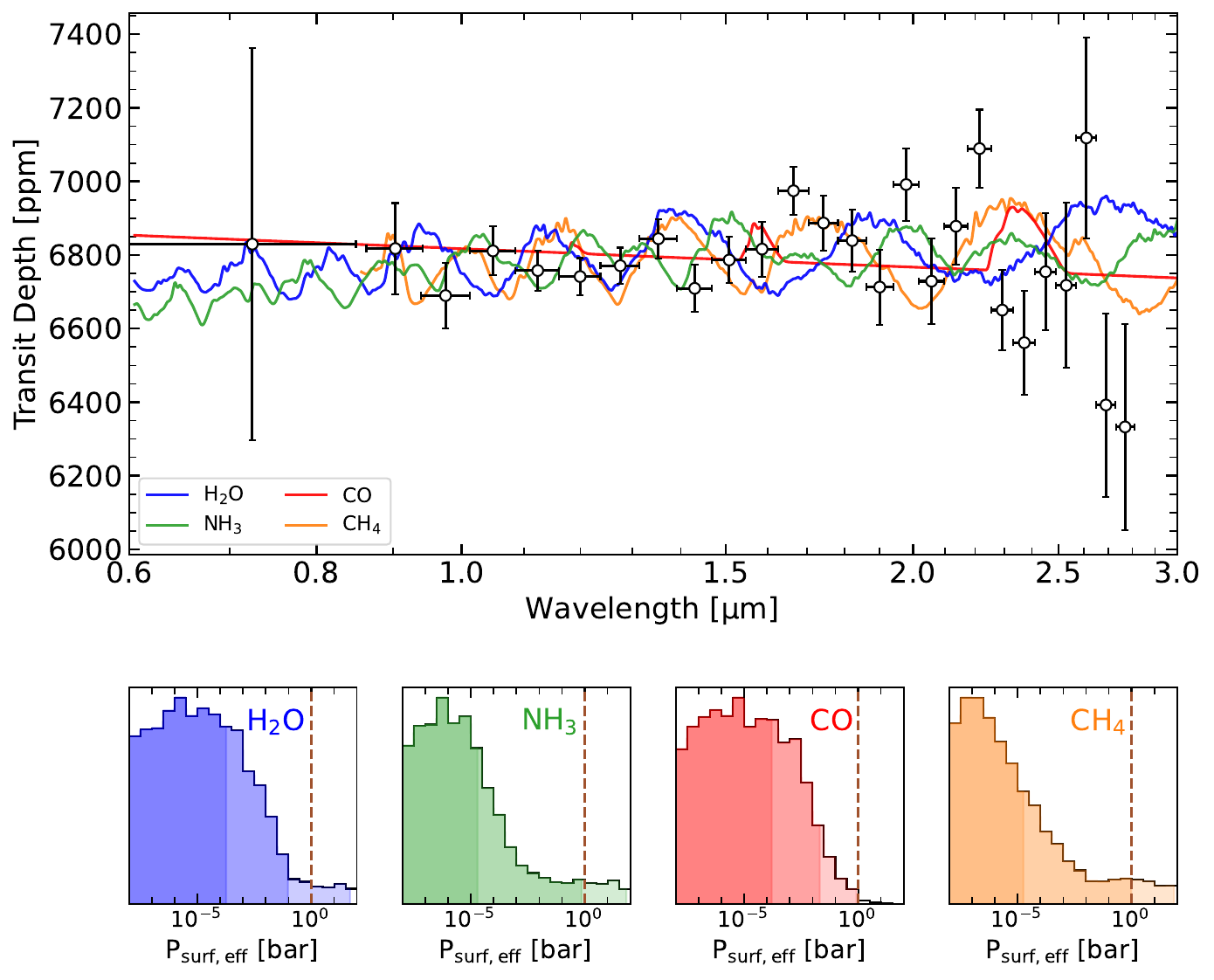}
    \caption{Constraints on high MMW atmosphere compositions for TRAPPIST-1\,c. 
    \textit{Top panel:} Combined visit 1+2 spectrum of TRAPPIST-1\,c corrected by the best-fitting stellar contamination model (black points, see text), and atmosphere models for deep, 100\,bar atmospheres composed of 100\% H$_2$O (blue), NH$_3$ (green), CO (red), and CH$_4$ (orange). 
    \textit{Bottom panels:} Marginalized posterior distributions on the effective surface pressure for different atmosphere compositions (labelled, with colours corresponding to the models in the top panel) The shaded areas correspond to the 1, 2, and 3$\sigma$ upper limits obtained on the effective surface pressure in each composition scenario. When marginalizing over the impact of stellar contamination, the transmission spectrum disfavors thick, $>$1\,bar atmospheres for all molecules at the $2\sigma$-level.
    \label{fig:atmosphere_molbymol}}
\end{figure*}

We also calculate forward models for 1 and 100$\times$ solar metallicity, H$_2$/He-dominated atmospheres with varying cloud-top pressures (Figure \ref{fig:met_psurf}; left panel), and models for several high MMW compositions (Figure \ref{fig:atmosphere_molbymol}; top panel), to put our retrieval results in context. For each model, the temperature-pressure profile is assumed to be isothermal at the equilibrium temperature of TRAPPIST-1\,c.

For each set of retrievals we account for the impacts of stellar contamination, thereby marginalizing over the interplay between unocculted stellar heterogeneities and planetary absorption features. In addition to the planetary atmosphere properties, Motivated by the results in Section~\ref{sec: TLSE Modelling}, we assume two heterogeneity populations: spots (cooler than the photosphere), and faculae (hotter than the photosphere). For each heterogeneity component, we fit a covering fraction (with a wide uniform prior of 0 to 50\%) and a temperature contrast to the photosphere (following the uniform priors of the TLS-only fit). We also fit for the photosphere temperature itself, with a Gaussian prior based on the value reported by \citet{agol_refining_2021}. 

We allow the photosphere $\log g$ to vary within the range $[2.5,5.5]$ as lower $\log g$ values are required to obtain a good match of the PHOENIX models to the out-of-transit spectrum of TRAPPIST-1 (see e.g., \citealp{lim_atmospheric_2023} and Section \ref{sec: out of transit}). Given that the TLS-only fit did not provide evidence for a difference in $\log g$ between the photosphere and heterogeneity components, we fix them to the same value for each model evaluation. Finally, we fit for an offset between the visit 1 and visit 2 spectra, motivated by the \texttt{stctm} retrieval results. 

\begin{figure*}
    \centering
    \includegraphics[width=0.9\linewidth]{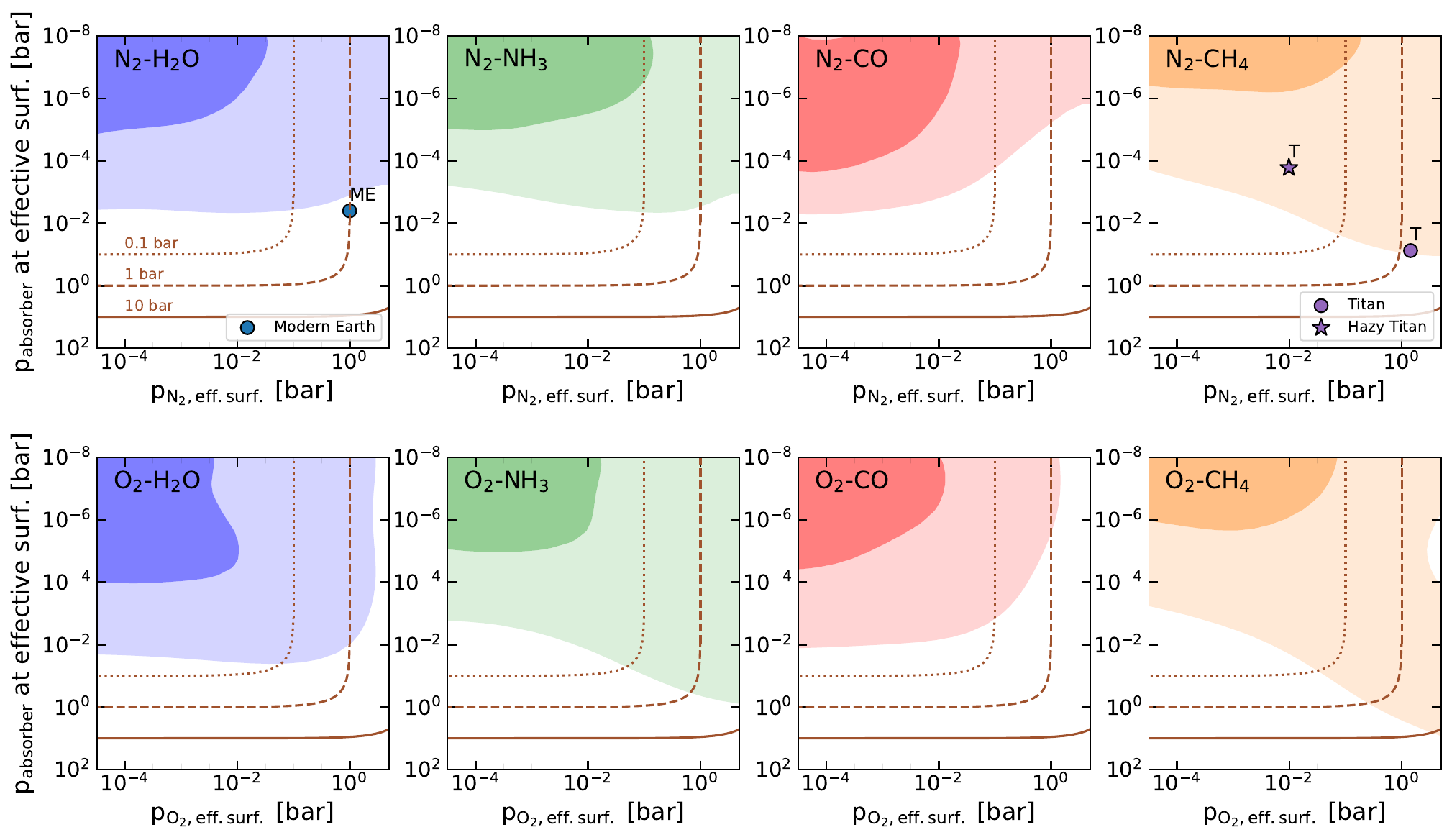}
    \caption{Constraints on N$_2$- and O$_2$-dominated atmospheres from the transmission spectrum of TRAPPIST-1\,c. 
    \textit{Top panels:} Two-dimensional posterior distribution of the partial pressure of the considered absorber at the effective surface pressure (left to right: H$_2$O, NH$_3$, CO, CH$_4$) as a function of the partial pressure of the background gas N$_2$ at the effective surface probed by the transmission spectrum. The 1 and 2$\sigma$ contours are filled with different colour shadings. The parts of the parameter space corresponding to effective surface pressures of 0.1, 1, and 10\,bar are outlined by the dotted, dashed, and solid sienna lines. We also illustrate where the amounts of methane in the atmosphere of Titan (with or without aerosols, purple markers; \citealp{charnay_titan_2014}) and of water in the atmosphere of Modern Earth (blue marker; \citealp{kasai_h2o_2011,kelsey_atmospheric_2022}) would lie relative to our posterior distributions.
    \textit{Bottom panels:} Same as the top panels, but for the retrievals with O$_2$ rather than N$_2$ as the background gas.}
    \label{fig:partialps}
\end{figure*}

Given the consistency between the stellar heterogeneity parameters inferred from both the TLS and stellar spectrum fits, we assume that the heterogeneity components for both visits are identical in our joint atmosphere fits. However, we note that we also performed one set of atmosphere+stellar contamination retrievals where we fit for visit-specific stellar heterogeneity properties, and found that this does not affect our inferences. We show the results for H$_2$/He-dominated atmospheres in Figure~\ref{fig:met_psurf}, and high-MMW atmospheres in Figures~\ref{fig:atmosphere_molbymol} and \ref{fig:partialps}. 

\section{Discussion} 
\label{sec: Discussion}

\subsection{Inferences of the Stellar Photosphere}
For the out-of-transit stellar modelling using both grids and visits, the photosphere temperature is generally higher than, but still broadly consistent with literature values \citep[e.g.,][]{agol_refining_2021, davoudi_updated_2024}. The retrieved log gravities ($\sim$3.8 for PHOENIX and $\sim$4.2 for SPHINX) are both significantly lower than the $\sim$5.2 found by \citet{agol_refining_2021} and \citet{davoudi_updated_2024}, through this discrepancy was also noted in the TRAPPIST-1 spectra obtained with NIRISS/SOSS by \citet{lim_atmospheric_2023} in their observations of planet b.

Our results from both grids are broadly consistent with the TLS retrievals (e.g., Figures~\ref{fig:stellar_contamination_results}, \ref{fig:stellar_spectrum_results}; Table~\ref{tab: Stellar Parameters}). For SPHINX, a two-component model including a photosphere and spots is preferred at $\sim$4$\sigma$ (using the \citet{benneke_how_2013} Bayesian evidence to $\sigma$-level conversion) over a homogeneous photosphere model and at $\sim$2$\sigma$ over a model including both spots and faculae for both visits. Like in the TLS retrievals, when faculae are included, the fits push the covering fraction to zero in order to minimize their contribution. Both the spot covering fraction and temperature are in excellent agreement with the TLS results. The spot temperatures are also consistent with those inferred by previous studies \citep[e.g.,][]{wakeford_disentangling_2019, garcia_hst_2022, lim_atmospheric_2023, davoudi_updated_2024}. For the PHOENIX grid, the results are also in excellent agreement with both the SPHINX grid and the TLS fit results (Table~\ref{tab: Stellar Parameters}). In this case, the three-component fit is preferred (though the significance is low; $\sim$1.8$\sigma$). The faculae covering fraction is also only $\sim$3\%, and therefore essentially negligible. For all grids tested, we find that the gravity of the heterogeneities is always consistent with that of the photosphere.

We note though that the overall quality of the fits are unsatisfactory, with $\chi^2_\nu$ values generally $>$350. The short wavelengths ($\lesssim$1.1\,µm) are especially badly fit by the PHOENIX grid in particular. In fact, the PHOENIX grid struggled to fit both the short and long wavelengths simultaneously, routinely overestimating the flux $>$1.7\,µm and finding photosphere temperatures that were highly inconsistent with literature values. We therefore restricted the models to wavelengths $>$1.1\,µm for the PHOENIX grid where the models were better matches to the data. Performing the same test for the SPHINX grid does not alter the retrieved parameter distributions.

Another caveat is that the high resolution of NIRSpec/G395H ($\sim$2700) allowed \citet{moran_high_2023} to fit to individual stellar lines. However, in our case, the medium resolution of NIRISS/SOSS ($\sim$700) does not provide this same capability (see the inset in Figure~\ref{fig:stellar_spectrum_results}), and our fits are more sensitive to the broadband structures in the spectrum. In all, though we confirm the results of the in-transit TLS fits, our stellar spectrum analysis demonstrates the clear need for improved stellar models for ultra-cool stars. Moreover, we posit that since the in-transit analysis relies on \textit{ratios} of stellar models instead of \textit{absolute fits}, it may be more robust to the aforementioned model inaccuracies than direct fits to stellar spectra --- something that should be explored further.

\subsection{Cloud-Free and Thick Hydrogen-Rich Atmospheres Are Ruled Out}

We find that, similarly to the closer-in planet TRAPPIST-1\,b \citep{lim_atmospheric_2023}, we can confidently rule out thick or cloud-free hydrogen-dominated atmospheres, with a 3-$\sigma$ limit of $<$10\,mbar on the effective surface pressure of a H$_2$/He-dominated atmosphere atop TRAPPIST-1\,c (Figure \ref{fig:met_psurf}). The thin hydrogen-rich atmosphere scenarios allowed by the transmission spectrum are, moreover, unlikely from an atmospheric evolution standpoint --- TRAPPIST-1\,c was exposed to large cumulative amounts of high-energy XUV irradiation from its host star, sufficient to strip away any primordial atmosphere \citep{hori_trappist-1_2020,turbet_review_2020} even when considering hydrogen outgassing \citep{hu_narrow_2023}. The transmission spectrum still allows within the 3-$\sigma$ confidence contours for thick, high-metallicity ($>$100$\times$ solar) atmospheres, which may be more resilient against escape due to their higher MMWs.

\subsection{Limits on High-Mean Molecular Weight Atmospheres}

We, therefore, turn to evaluate the range of plausible high-metallicity atmosphere scenarios for TRAPPIST-1\,c, and focus on the detectability of H$_2$O-, NH$_3$-, CO- and CH$_4$ in high-MMW atmospheres, as these infrared absorbers have sufficient opacity within the NIRISS/SOSS wavelength range to enable meaningful constraints. 

As expected from the high-metallicity tail of the surface pressure-metallicity distribution (Figure \ref{fig:met_psurf}), the densest atmosphere cases we tested do not provide strong upper limits on the effective surface pressure. We find that deep, 100\% H$_2$O, NH$_3$, and CH$_4$ atmospheres remain compatible with our spectrum at the 3$\sigma$ level, although $>$1\,bar atmospheres are disfavored at the 2$\sigma$ level (Figure \ref{fig:atmosphere_molbymol}). In the 100\% CO case, we exclude even a 1-bar atmosphere at the 3$\sigma$ level. However, we caution that this result mostly stems from the fact that CO has minimal opacity in the SOSS waveband from 2--2.5\,µm, and a pure-CO atmosphere would elsewhere yield a perfectly flat spectrum. Therefore, this example only rules out 100\% CO atmospheres, but likely not cases where a CO-rich atmosphere is mixed with small amounts of another infrared absorber.

We then turn to models that allow for lower atmospheric MMWs, where the infrared absorber can be a trace species rather than a dominant component --- representative of the atmospheres of some solar system planets. In this scenario, the background gas is N$_2$ or O$_2$, with only one of the aforementioned species as the active infrared absorber (Figure \ref{fig:partialps}). We compare these limits, where available, with measured compositions of solar system bodies. In particular, we highlight in Figure~\ref{fig:partialps} the amount of H$_2$O in Earth's N$_2$-dominated atmosphere (using 0.4\% as the mean mole fraction over the entire 1-bar atmosphere, although it can reach about one percent at sea level; \citealp{kasai_h2o_2011,kelsey_atmospheric_2022}), or the amount of CH$_4$ in Titan's atmosphere, assuming either a cloud-free Titan scenario (where one can see down to the surface) with 5\% CH$_4$, or a hazy scenario accounting for the fact that hydrocarbon haze opacity raises the effective surface pressure to $\sim$10\,mbar, and with a factor of three less methane in the upper atmosphere (see e.g. \citealp{charnay_titan_2014}). 

We find that similar amounts of water to what is observed in Earth's atmosphere are disfavored at the 2$\sigma$ level for TRAPPIST-1\,c, while the transmission spectrum alone is not sensitive enough to the CH$_4$ abundance to offer meaningful constraints on Titan-like atmospheres. Although our observations cannot rule out atmospheres where NH$_3$ or CH$_4$ make up the entire volatile budget (or are trace species in N$_2$- or O$_2$-dominated atmospheres), the presence of methane and ammonia is unlikely on TRAPPIST-1\,c from a modelling standpoint because of its vulnerability to destruction via photochemistry \citep{turbet_modeling_2018}. 

Since O$_2$ is a heavier molecule than N$_2$, the O$_2$ background scenarios generally correspond to more compact, higher-MMW atmospheres. Therefore, although the results are qualitatively similar to the N$_2$-background constraints, our transmission spectrum allows for larger amounts of the tested molecular absorbers in an O$_2$-dominated atmosphere. Further, because of the lack of sensitivity to CO$_2$ absorption with NIRISS/SOSS spectroscopy, we are not able to rule on any of the possible scenarios presented by \citet{zieba_no_2023} based on their MIRI eclipse observations of this planet. The photometric eclipse depth of 421$\pm$94\,ppm at 15\,µm was broadly consistent (better than 2$\sigma$) with a 0.01\,bar pure CO$_2$ atmosphere, or trace amounts of CO$_2$ in a O$_2$-dominated atmospheres with a surface pressure less than $\sim$10\,bar. We find that such an O$_2$-CO$_2$ atmosphere would give rise in transmission to a $\sim$40\,ppm CO$_2$ absorption feature at 2.7\,µm (at the red edge of the SOSS detector), and a larger $\sim$90\,ppm absorption band at 4.3\,µm. As demonstrated by \citet{taylor_awesome_2023}, CO$_2$ cannot be reliably probed at 2.7\,µm with SOSS alone, even for a hot-Jupiter, whereas the 4.3\,µm feature could potentially be detectable with $\sim$five transits with NIRSpec/G395M or PRISM. 

\subsection{Implications for the Presence of an Atmosphere on TRAPPIST-1c}

Although it could still be explained by some high metallicity scenarios, our non-detection of any significant atmospheric signature in the transmission spectrum of TRAPPIST-1 c is also in line with modeling predictions that TRAPPIST-1\,c could be a bare rock, as any substantial atmosphere is expected to be stripped over the 5--9\,Gyr lifetime of its host star \citep{burgasser_on_2017}, regardless of its initial volatile content \citep{krissansen-totton_implications_2023}.

Even if they remain tentative (at the 2$\sigma$ level), our atmospheric retrieval results disfavor thick $>$1\,bar atmospheres, even for high MMW compositions. The thin atmosphere scenarios that remain consistent with our observations given their smaller spectral features are however unlikely to be retained on TRAPPIST-1\,c from a theoretical perspective. Even if Jeans escape (generally less efficient than blow-off) is expected to dominate for high-MMW atmospheres atop TRAPPIST-1\,c, the escape rates even for resilient N$_2$-CO$_2$ atmospheres range from one Earth atmosphere per Myr to one Earth atmosphere mass per Gyr \citep{van_looveren_2024} at the present-day irradiation of this planet. The implications are even stricter when considering the fact that any atmosphere on TRAPPIST-1\,c also would have had to withstand the early XUV-active stages of the late M dwarf's evolution \citep{bolmont_water_2017}, and its position inside of the runaway greenhouse limit \citep{turbet_review_2020} indicates that water could not have been efficiently shielded from escape via condensation in a liquid water ocean, even if this limit is pushed to closer host star-planet distances around M dwarfs in 3D models \citep{turbet_water_2023}.

\subsection{Impacts of Stellar Contamination on the Transmission Spectra}
\label{sec: TLS Limitations}

Our two-transit study of TRAPPIST-1\,c aptly summarizes the promise and peril of transmission observations of rocky planets around M dwarf host stars. There are now a plethora of studies that have shown that stellar contamination is a true bottleneck in the attempt to detect atmospheres around rocky, M dwarf planets via transmission spectroscopy \citep[e.g.,][]{zhang_near_2018, wakeford_disentangling_2019, may_double_2023, moran_high_2023, lim_atmospheric_2023, cadieux_transmission_2024, rackham_toward_2024}. In most cases, different manifestations of the TLSE (i.e., different temperature distributions, covering fractions, etc.~of stellar spots and faculae) give rise to obviously different transmission spectra between visits, allowing for a clear inference of the impacts of the TLSE, even without undertaking in-depth modelling. Consistency in the location and shapes of spectral features between visits has been argued to lend credence to an interpretation favouring a planetary atmosphere origin \citep[e.g.][]{may_double_2023}. 

However, despite the high level of consistency between the transmission spectra ($\sim$1.1-$\sigma$ on average), our retrieval analysis still finds that both visits can be entirely explained by stellar contamination. As shown in Figure~\ref{fig:TLS_inferences}, and summarized in Table~\ref{tab: Stellar Parameters}, we find very similar stellar heterogeneity parameters (spot/faculae temperatures and covering fractions) for both visits. We also find both visits to be spot-dominated, with minimal contributions from faculae, whereas \citet{lim_atmospheric_2023} have one spot-dominated, and one facula-dominated visit in their two-transit analysis of TRAPPIST-1\,b. 

The consistency between visits would perhaps be unsurprising if they were separated by a very small period of time, which would therefore probe similar spot and facula distributions on the host star. However, our two visits are separated by $\sim$367\,d, significantly longer than the 3.3\,d stellar rotation period \citep{luger_seven-planet_2017}. When phase-folded to the stellar rotation period, this corresponds to a phase difference of $\sim$0.21 --- similar to the observations of \citet{lim_atmospheric_2023}, who, as mentioned above, found no such consistency between their two visits. Moreover, both of our visits are at comparable stellar phases to those of \citet{lim_atmospheric_2023}. This highlights the fact that the distributions of spots and faculae in the host star's photosphere are entirely stochastic, and therefore TLSE models are non-predictive of future stellar contamination. 

Irrespective of whether we perform TLS-only or joint TLS-atmosphere retrievals, we still infer consistent stellar contamination parameters (Figure \ref{fig:TLS_inferences}). This is consistent with our finding that the retrieval does not pick up on any ``true'' atmosphere signature that would warrant changes in the inferred TLS parameters.

Finally, we evaluated the limiting effect of having to marginalize over stellar contamination on our sensitivity to thin 100\% H$_2$O atmospheres. This choice was motivated by the fact that water is the molecule most affected by this marginalization, given that its features coincide with prominent unocculted stellar spot signatures, but refer the reader to Piaulet-Ghorayeb et al. (in prep) for a more detailed exploration of the impact of TLSE marginalization on other infrared absorbers. We performed a retrieval on the TLS-corrected spectra from both visits (see Section~\ref{sec: in transit}), using the exact same setup as the joint atmosphere + TLS retrieval except that only the atmosphere properties are retrieved.

We find that our sensitivity drops by about two orders of magnitude through accounting for stellar contamination. As shown in Figure~\ref{fig:TLS_marg}, without stellar contamination we could rule out at the 2$\sigma$ (3$\sigma$) level 100\% H$_2$O atmospheres with surface pressures down to $\sim$1\,mbar (1.6\,bar). However, when marginalizing over the TLSE, we can only rule out at the 2$\sigma$ (3$\sigma$) level cases with surface pressures of less than $\sim$0.1\,bar (45\,bar). Therefore, we are in a regime where the unknown TLS contribution to the spectrum, rather than the precision of the measured transit depths themselves, is the limiting factor in our sensitivity to the presence of high-MMW atmospheres. Furthermore, this effect will be even more problematic in other datasets where one needs to marginalize over different stellar contamination realizations for each visit in joint retrievals (e.g., Piaulet-Ghorayeb et al., in prep).

\section{Conclusions}
\label{sec: Conclusions}

Here, we presented the first JWST transmission spectrum of the second innermost planet in the TRAPPIST-1 system; TRAPPIST-1\,c. Our 0.6 -- 2.85\,µm NIRISS/SOSS spectrum reveals strong contamination by the TLSE in both visits. However, unlike in previous multi-visit analyses of rocky planets around M dwarfs  \citep[e.g.,][]{may_double_2023, moran_high_2023, cadieux_transmission_2024}, including TRAPPIST-1\,b \citep{lim_atmospheric_2023}, we find consistent stellar heterogeneity properties between our two visits. Via an analysis of the in-transit data (i.e., the TLSE) and direct modelling of the out-of-transit stellar spectrum, we find the stellar photosphere to be dominated by spots $\sim$300\,K cooler than the ambient photosphere temperature and with a covering fraction of $\sim$25\%. Faculae have negligible contributions to both visits, irrespective of the analysis technique or choice of stellar models. This results in nearly identical manifestations of the TLSE and thus transmission spectra. Our results support the need for joint modelling of the TLSE and atmosphere properties to achieve robust constraints on rocky, M dwarf planets in the era of JWST. Importantly, further stellar model fidelity is critically required to accurately model out the impact of the TLSE and reach the sensitivity to detect or rule out thin atmospheres on small planets via their transit spectra.

When accounting for stellar contamination in our atmosphere analysis, we find that thick, clear hydrogen-dominated compositions are ruled out at better than 3-$\sigma$ significance. The level of precision we reach does not enable us to rule out the cases with the highest MMW, where the atmosphere would be composed of 100\% of a heavy volatile species such as H$_2$O, NH$_3$, or CH$_4$, but even these scenarios are disfavored for $>$1\,bar atmospheres at the $2\sigma$-level due to the flatness of the TLSE-corrected spectrum. The sensitivity increases to partial pressures of about $\lesssim$10\,mbar at 2$\sigma$ for the same species in atmospheres with N$_2$ or O$_2$ as the background gas. 

Regardless, as shown by \citet{krissansen-totton_implications_2023}, even if TRAPPIST-1\,c is completely atmosphere-less, this does not imply a similar fate for the outer, habitable-zone planets. Our work, though, does highlight the fact that we cannot simply rely on consistency of transmission spectra between different visits to infer the presence of a planetary atmosphere and rule out stellar contamination.

\begin{acknowledgments}
M.R.\ would like to acknowledge funding from the Natural Sciences and Research Council of Canada (NSERC), as well as from the Fonds de Recherche du Quebec Nature et Technologies (FRQNT). 
C.P.-G acknowledges support from the NSERC Vanier scholarship, and the Trottier Family Foundation. C.P.-G also acknowledges support from the E. Margaret Burbidge Prize Postdoctoral Fellowship from the Brinson Foundation. 
J.T was supported by the Glasstone Benefaction, University of Oxford (Violette and Samuel Glasstone Research Fellowships in Science 2024). 
A.L’H. acknowledges support from the FRQNT under file \#349961. 
O.L. acknowledges financial support from the FRQNT. The authors acknowledge the financial support of the FRQNT through the {\it Centre de recherche en astrophysique du Québec} as well as the support from the Trottier Family Foundation and the Trottier Institute for Research on Exoplanets.
This work is based on observations made with the NASA/ESA/CSA JWST. The data were obtained from the Mikulski Archive for Space Telescopes at the Space Telescope Science Institute, which is operated by the Association of Universities for Research in Astronomy, Inc., under NASA contract NAS 5-03127 for JWST. The specific observations analyzed can be accessed via~\dataset[10.17909/ys1r-b952]{10.17909/ys1r-b952}. This research has made use of the NASA Exoplanet Archive, which is operated by the California Institute of Technology, under contract with the National Aeronautics and Space Administration under the Exoplanet Exploration Program.
\end{acknowledgments}

\vspace{5mm}
\facilities{JWST(NIRISS), Exoplanet Archive}

\software{\texttt{astropy} \citep{astropy:2013, astropy:2018}, 
\texttt{celerite} \citep{foreman-mackey_fast_2017},
\texttt{exoTEDRF} \citep{radica_exotedrf_2024},
\texttt{exoUPRF} \citep{radica_exouprf_2024},
\texttt{ipython} \citep{PER-GRA:2007},
\texttt{juliet} \citep{espinoza_juliet_2019},
\texttt{jwst} \citep{bushouse_howard_2022_7038885},
\texttt{matplotlib} \citep{Hunter:2007},
\texttt{numpy} \citep{harris2020array},
\texttt{pymultinest} \citep{buchner_statistical_2016},
\texttt{scipy} \citep{2020SciPy-NMeth}
}

\appendix

\section{\texttt{NAMELESS} Reduction and ExoTEP Light Curve Fitting}
\label{sec: Additional Reductions}

We reduce our two JWST NIRISS/SOSS visits of TRAPPIST-1c, starting from the uncalibrated raw data up to the extraction of the stellar spectra, using the \texttt{NAMELESS} pipeline \citep{coulombe_broadband_2023,feinstein_early_2023}. We first go through the Stages 1 and 2 steps of the STScI \texttt{jwst} pipeline (v1.12.5) \citep{bushouse_howard_2022_7038885}, which include super-bias subtraction, reference pixel correction, non-linearity correction, ramp-fitting, and flat-fielding. We then proceed with a sequence of custom steps to correct for remaining sources of noise such as bad pixels, non-uniform background, cosmic-rays, and 1/$f$ following the methods described in \citet{benneke_jwst_2024}. We flag bad pixels by looking for pixels whose second derivative are significant ($>$7$\sigma$) outliers in the spatial domain. The counts of all flagged pixels are then corrected using the bicubic interpolation method of the \texttt{scipy.interpolate.griddata} function. We correct for the non-uniform background by individually scaling the two regions of the STScI model background\footnote{\url{https://jwst-docs.stsci.edu/}} that are separated by the sudden jump in background flux situated around spectral pixel x$\sim$700. Any remaining cosmic rays are corrected by clipping any count that is more than four standard deviations away from the running median for all pixels. As for the 1/$f$ noise correction, we follow the same method described in \citet{coulombe_broadband_2023} and \citet{benneke_jwst_2024} where we scale individually each column of the first and second spectral orders considering only pixels that are within a 30-pixel distance from the center of the traces to compute the 1/$f$ noise. We then extract the stellar spectra from both spectral orders using a simple box aperture with a width of 36 pixels. 

\begin{figure*}
	\centering
	\includegraphics[width=0.9\textwidth]{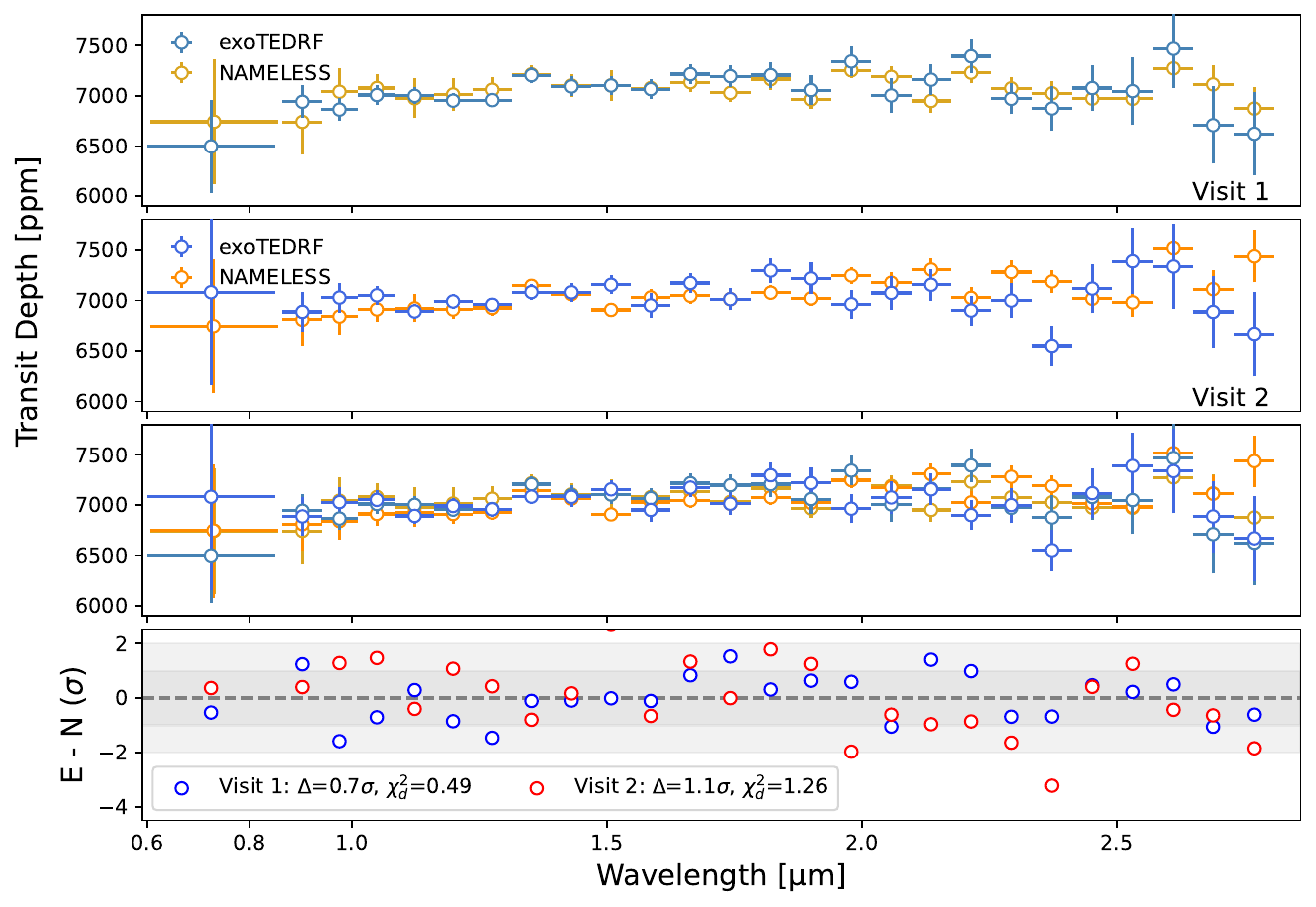}
    \caption{Comparison of transmission spectra from the \texttt{exoTEDRF} and \texttt{NAMELESS} pipelines. 
    \emph{Top panel:} Spectrum comparison for visit 1.
    \emph{Second panel:} Same as top but for visit 2.
    \emph{Third panel:} All four spectra from both pipelines and for both visits overplotted to show consistency. 
    \emph{Bottom panel:} Differences between spectra from each pipeline for visit 1 (blue) and visit 2 (red) divided by the error bar on each transit depth. The \texttt{exoTEDRF} and \texttt{NAMELESS} spectra are consistent to within 0.7$\sigma$ on average for visit 1 and 1.1$\sigma$ for visit 2. Also quoted are comparisons between the two reductions using a $\chi^2$-per-data-point metric.
    \label{fig:CompareSpectra}}
\end{figure*}

We perform the broadband and spectroscopic fits of the light curves with the ExoTEP framework \citep{benneke_sub-neptune_2019,benneke_water_2019}. For our fit of the white-light curve, which is obtained by summing all wavelengths of the first spectral order (0.85--2.85\,$\mu$m), we keep free the mid-transit time ($T_0$, considering uniform priors spanning the time of the observations for each visit), planet-to-star radius ratio ($R_p/R_*$, $\mathcal{U}$[0.01,0.2]), semi-major axis ($a/R_*$, $\mathcal{U}$[15,45]), and the impact parameter ($b$, $\mathcal{U}$[0,1]). We consider the quadratic limb-darkening law and fit directly for $u_1$ and $u_2$ assuming large uniform priors ($\mathcal{U}$[-3,3]) following \citet{Coulombe2024_biases}. For the first visit, we consider a quadratic trend for the systematics model. We also include in the fit a Matérn 3/2 Gaussian Process using the \texttt{celerite} python \citep{foreman-mackey_fast_2017}  to remove higher-frequency trends that are not corrected by our systematics model. The timescale and amplitude of the GP model are kept free assuming large uninformative uniform priors in log-space. As for the second visit, we cut the first 2.1 hours of the TSO and consider solely a linear slope for the systematics model. The transit light curves are then modelled in ExoTEP using the \texttt{batman} \citep{kreidberg_batman_2015} python package, and we explore the parameter space using and the Markov chain Monte Carlo (MCMC) sampler \texttt{emcee} \citep{foreman-mackey_emcee_2013}. We run the MCMC for 10,000 steps and discard the first 6,000 steps as burn-in. The same procedure is repeated for the spectroscopic light curves (assuming the same systematics model and GP for each visit), where we have fixed the orbital parameters of TRAPPIST-1\,c to the best-fit values from the white-light curve of the first visit ($a/R_*$ = 28.55, $b$ = 0.112). The light curves are fit using 100 and 30 equal-pixel bins for the first and second spectral orders, respectively. The resulting spectra, binned to a fixed resolving power of $R = 25$, are shown in Figure~\ref{fig:CompareSpectra}.

\section{Additional Figure \& Tables}
\label{sec: Additional Tables}

Table~\ref{tab: WLC Parameters} contains the best-fitting white light curve parameters for both NIRISS/SOSS visits. The orbital parameters ($a/R_*$, $inc$) are consistent with each other, and with previous literature \citep[e.g.,][]{agol_refining_2021}.

\begin{table}
\centering
\caption{Best-fitting \texttt{exoTEDRF} white light transit parameters}
\label{tab: WLC Parameters}
\begin{threeparttable}
    \begin{tabular}{c|cccc}
        \hline
        \hline
        \multirow{2}{*}{Parameter} & Prior Range & \multicolumn{3}{c}{Value} \\
         & & Visit 1 & Visit 2 & Weighted Average \\
        \hline
        T0 [BJD] & $\mathcal{U}$[T0$\pm$2\,hr] & 2459881.401528$^{+0.000033}_{-0.000033}$ & 2460249.515031$^{+0.000042}_{-0.000041}$ & $-$ \\
        $\rm R_p/R_{*,O1}$ & $\mathcal{U}$[0.01, 0.9] & 0.08346$^{+0.00092}_{-0.00097}$ & 0.08344$^{+0.00163}_{-0.00141}$ & $0.08345\pm0.00112$ \\
        $\rm R_p/R_{*,O2}$ & $\mathcal{U}$[0.01, 0.9] & 0.08364$^{+0.00260}_{-0.00265}$ & 0.08415$^{+0.00881}_{-0.00970}$ & $0.08379\pm0.00472$ \\
        $\rm inc$ & $\mathcal{U}$[70, 90] & 89.73$^{+0.17}_{-0.22}$ & 89.38$^{+0.32}_{-0.26}$ & $89.64\pm0.24$ \\
        $\rm a/R_*$ & $\mathcal{U}$[5, 50] & 28.35$^{+0.27}_{-0.51}$ & 28.18$^{+0.85}_{-1.05}$ & $28.3\pm0.54$ \\
        $\rm u_{1,O1}$ & $\mathcal{U}$[-1, 1] & 0.405$^{+0.100}_{-0.105}$ & 0.283$^{+0.197}_{-0.212}$ & $0.344\pm0.134$ \\
        $\rm u_{2,O1}$ & $\mathcal{U}$[-1, 1] & 0.168$^{+0.187}_{-0.174}$ & 0.346$^{+0.299}_{-0.289}$ & $0.257\pm0.217$ \\
        $\rm u_{1,O2}$ & $\mathcal{U}$[-1, 1] & 0.671$^{+0.229}_{-0.348}$ & 0.302$^{+0.442}_{-0.510}$ & $0.487\pm0.344$ \\
        $\rm u_{2,O2}$ & $\mathcal{U}$[-1, 1] & $-$0.098$^{+0.509}_{-0.402}$ & 0.105$^{+0.567}_{-0.561}$ & $0.003\pm0.477$ \\
        \hline
    \end{tabular}
    \begin{tablenotes}
        \small
        \item \textbf{Note}: For each visit, $\rm R_p/R_*$ and the limb darkening parameters ($\rm u_1$ and $\rm u_2$)  were fit separately for each order. All other parameters were jointly fit to both.
    \end{tablenotes}
\end{threeparttable}
\end{table}

Table~\ref{tab: Stellar Parameters} contains the best-fitting parameters from the three-component stellar models via both the in-transit (TLS) and out-of-transit analyses.

\begin{table}
\centering
\caption{Retrieved parameters for the \texttt{stctm} retrievals performed on the transmission spectra (in-transit, TLS effect) assuming no atmosphere and two heterogeneity components, and from the three-component fit to the out-of-transit spectra using \texttt{StellarFit}.  We quote the median parameters and $\pm1$-$\sigma$ intervals, or 2-$\sigma$ limits when a parameter hits the edge of a prior.}
\label{tab: Stellar Parameters}
\begin{threeparttable}
    \begin{tabular}{c|ccccc}
        \hline
        \hline
        \multirow{2}{*}{Parameter} & \multirow{2}{*}{Model} & \multicolumn{2}{c}{In-Transit (TLS)} & \multicolumn{2}{c}{Out-of-Transit} \\
         & & Visit 1 & Visit 2 & Visit 1 & Visit 2 \\
        \hline
        \multirow{2}{*}{T$\rm _{eff}$ [K]} & PHOENIX & $2578^{+68}_{-63}$ & $2594^{+64}_{-59}$& $2591^{+18}_{-22}$ & $2595^{+19}_{-21}$ \\
         & SPHINX & & & $2597^{+5}_{-8}$ & $2596^{+5}_{-9}$ \\
        \rule{0pt}{4ex} \multirow{2}{*}{$\rm \log g_{phot}$} & PHOENIX & $4.5^{+0.7}_{-1.0}$ & $4.3^{+0.7}_{-0.8}$ & $3.6^{+0.1}_{-0.1}$ & $3.5^{+0.1}_{-0.1}$ \\
         & SPHINX & & & $4.2^{+0.1}_{-0.1}$ & $4.2^{+0.1}_{-0.1}$ \\
        \rule{0pt}{4ex} \multirow{2}{*}{$\rm T_{spot}$ [K]} & PHOENIX & $<2510$ & $<2523$ & $<$2437 & $<$2432 \\
         & SPHINX & & & $2286^{+55}_{-54}$ & $2284^{+74}_{-58}$ \\
        \rule{0pt}{4ex} \multirow{2}{*}{$\rm f_{spot}$} & PHOENIX & $0.19^{+0.14}_{-0.11}$ & $0.19^{+0.14}_{-0.11}$  & $0.17^{+0.06}_{-0.05}$ & $0.17^{+0.06}_{-0.05}$ \\
         & SPHINX & & & $0.24^{+0.03}_{-0.03}$ & $0.22^{+0.03}_{-0.02}$ \\
        \rule{0pt}{4ex} \multirow{2}{*}{$\rm \log g_{spot}$} & PHOENIX & $>2.6$ & $>2.7$ & $3.6^{+0.1}_{-0.1}$ & $3.6^{+0.1}_{-0.1}$ \\
         & SPHINX & & & $4.2^{+0.1}_{-0.1}$ & $4.2^{+0.1}_{-0.1}$ \\
        \rule{0pt}{4ex} \multirow{2}{*}{$\rm T_{fac}$ [K]} & PHOENIX & $2783^{+175}_{-102}$ & $2831^{+225}_{-117}$ & $3324^{+55}_{-85}$ & $3396^{+49}_{-71}$ \\
         & SPHINX & & & $<$3054 & $<$3133 \\
        \rule{0pt}{4ex} \multirow{2}{*}{$\rm f_{fac}$} & PHOENIX & $<0.39$& $<0.32$ & $0.03^{+0.01}_{-0.01}$ & $0.03^{+0.01}_{-0.01}$ \\
         & SPHINX & & & $<$0.07 & $<$0.05 \\
        \rule{0pt}{4ex} \multirow{2}{*}{$\rm \log g_{fac}$} & PHOENIX & $>2.6$ & $>2.7$  & $3.6^{+0.1}_{-0.2}$ & $3.6^{+0.2}_{-0.2}$ \\
         & SPHINX & & & $4.1^{+0.2}_{-0.2}$ & $4.3^{+0.2}_{-0.1}$\\
        \hline
    \end{tabular}
    \begin{tablenotes}
        \small
        \item \textbf{Note}: For the \texttt{stctm} fits, the values quoted for $T_\mathrm{spot}$ and $T_\mathrm{fac}$ are derived from the posterior samples of $\Delta T_\mathrm{spot}$, $\Delta T_\mathrm{fac}$, and $ T_\mathrm{phot}$, and the intervals reported for the heterogeneity $\log g$ are derived from the $\log g_\mathrm{phot}$ and $\Delta \log g_\mathrm{het}$ samples.
    \end{tablenotes}
\end{threeparttable}
\end{table}

Table~\ref{tab: spectrum} has the \texttt{exoTEDRF} transmission spectrum used in this work.

\begin{table}
\centering
\caption{\texttt{exoTEDRF} transmission spectrum of TRAPPIST-1\,c used in this work}
\label{tab: spectrum}
\begin{threeparttable}
    \begin{tabular}{cccc}
        \hline
        \hline
        Wavelength & Wavelength Error & Transit Depth (visit 1) & Transit Depth (visit 2) \\
        (µm) & (µm) & (ppm) & (ppm)  \\
        \hline
        0.72 & 0.12 & 6495.76 $\pm$ 463.23 & 7080.25 $\pm$ 918.63 \\
        0.90 & 0.04 & 6940.96 $\pm$ 162.46 & 6884.63 $\pm$ 193.99 \\
        0.98 & 0.04 & 6863.36 $\pm$ 112.06 & 7026.99 $\pm$ 147.35 \\
        1.05 & 0.04 & 7008.37 $\pm$ 98.91 & 7049.40 $\pm$ 95.48 \\
        1.12 & 0.04 & 6999.78 $\pm$ 83.35 & 6889.46 $\pm$ 77.15 \\
        1.20 & 0.04 & 6952.18 $\pm$ 71.99 & 6988.60 $\pm$ 74.63 \\
        1.28 & 0.04 & 6957.36 $\pm$ 70.95 & 6955.08 $\pm$ 72.35 \\
        1.35 & 0.04 & 7205.70 $\pm$ 81.75 & 7081.42 $\pm$ 78.11 \\
        1.43 & 0.04 & 7093.85 $\pm$ 98.13 & 7079.19 $\pm$ 95.84 \\
        1.51 & 0.04 & 7103.09 $\pm$ 94.84 & 7154.09 $\pm$ 93.00 \\
        1.59 & 0.04 & 7066.62 $\pm$ 97.78 & 6949.11 $\pm$ 120.23 \\
        1.66 & 0.04 & 7216.39 $\pm$ 96.84 & 7171.97 $\pm$ 95.48 \\
        1.74 & 0.04 & 7195.63 $\pm$ 109.50 & 7011.15 $\pm$ 110.87 \\
        1.82 & 0.04 & 7206.13 $\pm$ 127.63 & 7295.62 $\pm$ 123.62 \\
        1.90 & 0.04 & 7057.07 $\pm$ 145.74 & 7217.11 $\pm$ 158.78 \\
        1.98 & 0.04 & 7342.25 $\pm$ 148.19 & 6961.31 $\pm$ 144.52 \\
        2.06 & 0.04 & 7004.14 $\pm$ 175.93 & 7073.29 $\pm$ 167.36 \\
        2.14 & 0.04 & 7160.88 $\pm$ 151.13 & 7154.43 $\pm$ 157.82 \\
        2.21 & 0.04 & 7394.33 $\pm$ 166.09 & 6897.93 $\pm$ 147.24 \\
        2.29 & 0.04 & 6970.20 $\pm$ 153.40 & 6997.21 $\pm$ 172.76 \\
        2.37 & 0.04 & 6874.09 $\pm$ 223.18 & 6547.71 $\pm$ 198.38 \\
        2.45 & 0.04 & 7077.64 $\pm$ 225.76 & 7116.19 $\pm$ 245.70 \\
        2.53 & 0.04 & 7046.87 $\pm$ 340.02 & 7388.53 $\pm$ 327.30 \\
        2.61 & 0.04 & 7468.39 $\pm$ 389.34 & 7336.15 $\pm$ 420.91 \\
        2.69 & 0.04 & 6707.51 $\pm$ 385.19 & 6884.10 $\pm$ 356.58 \\
        2.77 & 0.04 & 6621.37 $\pm$ 414.79 & 6666.19 $\pm$ 417.26 \\
        \hline
    \end{tabular}
\end{threeparttable}
\end{table}

\begin{figure*}
    \centering
    \includegraphics[width=0.7\linewidth]{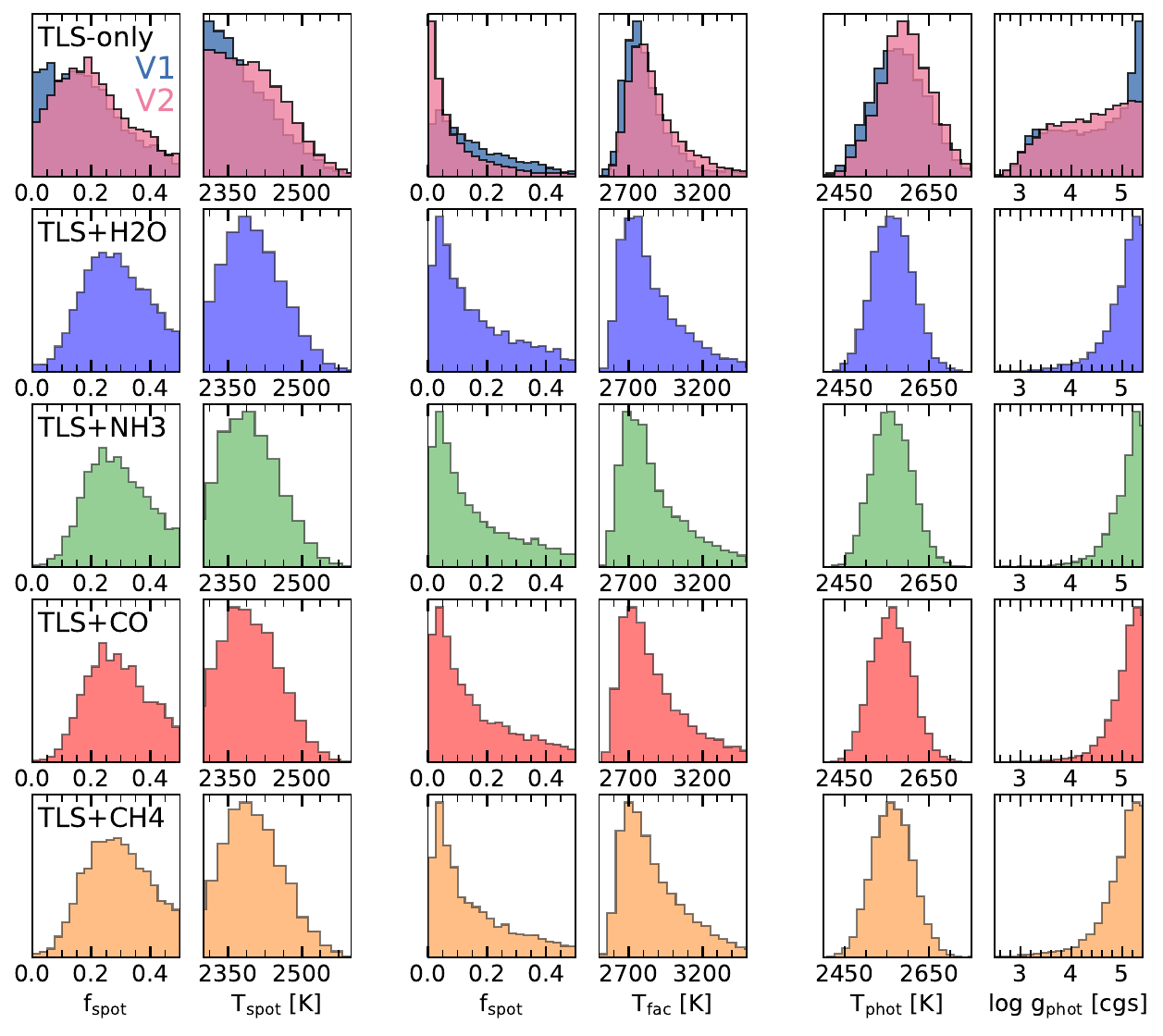}
    \caption{Posterior distributions of the inferred stellar surface properties from the TLS-only \texttt{stctm} retrieval (top row; visit 1 in blue, visit 2 in pink), and from the joint stellar heterogeneity + planetary atmosphere retrievals performed with SCARLET, for the 100\% H$_2$O, NH$_3$, CO, and CH$_4$ cases (susequent rows; labelled).
    \label{fig:TLS_inferences}}
\end{figure*}

We compare the posterior distributions of the spot, facula, and photosphere properties inferred from the TLS-only fit to the visit 1 and visit 2 spectra, and the joint atmosphere and stellar contamination retrievals (for the single-component atmosphere cases) in Figure \ref{fig:TLS_inferences}.

\begin{figure} 
	\centering
	\includegraphics[width=0.4\textwidth]{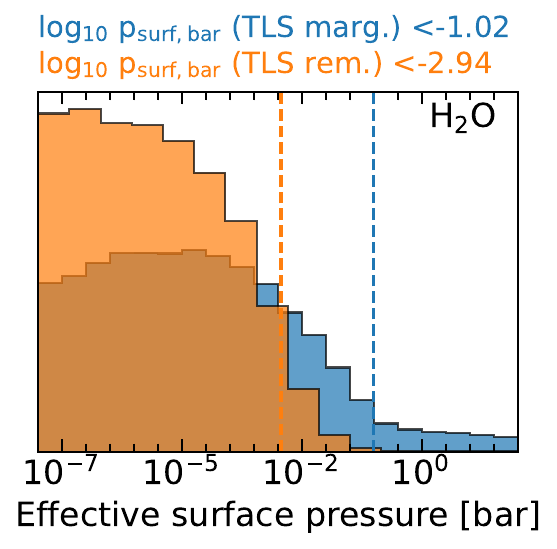}
    \caption{Impact of marginalizing over all the possible TLS realizations rather than retrieving directly on a ``TLS-corrected'' spectrum with the same error bars. The posterior distribution on the H$_2$O effective partial pressure (from the 100\% H$_2$O atmosphere retrieval) is shown for two cases: one where we perform a joint stellar contamination + planetary atmosphere retrieval on the observed spectrum (blue) and one where an atmosphere-only retrieval is performed on our best estimate of a TLS-free spectrum (orange). The 2$\sigma$ upper limits we obtain are quoted at the top of the figure, and shown as dashed vertical. Marginalizing over the TLSE worsens our sensitivity to 100\% H$_2$O atmospheres by roughly two orders of magnitude.
    \label{fig:TLS_marg}}
\end{figure}

Finally, we also present the results of our sensitivity study assessing the impact of marginalizing over the TLS effect on the atmosphere thicknesses that can be ruled out by our spectrum, in Figure~\ref{fig:TLS_marg}.

\bibliography{main}{}
\bibliographystyle{aasjournal}

\end{document}